\documentclass[letterpaper,journal]{IEEEtran}
\usepackage{amsmath,amsfonts,amssymb,amsthm}
\usepackage{algorithmic}
\usepackage{algorithm}
\usepackage{array}
\usepackage{textcomp}
\usepackage{stfloats}
\usepackage{url}
\usepackage{verbatim}
\usepackage{graphicx}
\usepackage[caption=false,font=footnotesize]{subfig}
\usepackage{cite}
\usepackage{bm}
\usepackage{mathrsfs}
\usepackage{soul,xcolor}

\newtheorem{theorem}{Theorem}

\allowdisplaybreaks[4]

\def\BibTeX{{\rm B\kern-.05em{\sc i\kern-.025em b}\kern-.08em
		T\kern-.1667em\lower.7ex\hbox{E}\kern-.125emX}}
\usepackage{balance}
\AtBeginDocument{%
  \renewenvironment{proof}[1][\IEEEproofname]{%
    \par\noindent{\itshape #1: }\ignorespaces
  }{%
    \hspace*{\fill}~\IEEEQED\par
  }%
}
\begin{document}
	\title{Microwave Linear Analog Computer (MiLAC) Enabled Wireless Powered MEC Networks}
	\author{Yixuan Li, Ji Wang, \IEEEmembership{Senior Member,~IEEE}, Wenwu Xie, Jiancheng An, \IEEEmembership{Senior Member,~IEEE},\\ Qimei Cui, \IEEEmembership{Senior Member,~IEEE}, and Liang Yang, \IEEEmembership{Senior Member,~IEEE}
		\thanks{\textit{(Corresponding author: Ji Wang.)}}
		\thanks{Yixuan Li and Ji Wang are with the Department of Electronics and Information Engineering, College of Physical Science and Technology, Central China Normal University, and are also with Academy of Frontier Interdisciplinary Research, Central China Normal University, Wuhan 430079, China (e-mail: yixuanli@mails.ccnu.edu.cn; jiwang@ccnu.edu.cn).
			
		Wenwu Xie is with the School of Information Science and
		Engineering, Hunan Institute of Science and Technology, Yueyang 414006, China (e-mail: gavinxie@hnist.edu.cn).
		
		Jiancheng An is with the School of Electronic Science
		and Engineering, University of Electronic Science and Technology of
		China, Chengdu, 611731, China (e-mail: jiancheng.an@uestc.edu.cn).
		
		Qimei Cui is with the National Engineering Research Center for
		Mobile Network Technologies, Beijing University of Posts and Telecommunications, Beijing, 100876, China,  and also with the
		Department of Broadband Communication, Pengcheng Laboratory, Shenzhen 518000, China (e-mail: cuiqimei@bupt.edu.cn).
		
		Liang Yang is with the College of Computer Science and Electronic
		Engineering, Hunan University, Changsha 410082, China, and also with School of Information Engineering, Changsha Medical University, Changsha 410219, China (e-mail: liangy@hnu.edu.cn).
	}}
	
	\maketitle
	
	\begin{abstract}
		Microwave linear analog computers (MiLACs) enable flexible analog beamforming with reduced radio-frequency (RF)-chain requirements. This paper investigates a wireless powered mobile edge computing (WP-MEC) network enabled by a hardware-efficient stem-connected MiLAC (SC-MiLAC) for dynamic wireless power transfer (WPT) and uplink time-division multiple access (TDMA). We prove that, under linear energy harvesting (EH) and negligible switching delays, a dynamic single-RF-chain SC-MiLAC can achieve the same WPT performance and minimum task completion latency as fully digital architectures. We further establish theoretical performance guarantees for dynamic SC-MiLAC under nonlinear EH and finite switching delays, characterizing the required beam configurations and the impact of switching delays on system performance.
		Considering nonlinear EH, hardware switching delays, partial task offloading, and limited edge computing resources, we formulate a joint beamforming and resource allocation problem to minimize the maximum task completion latency. To solve the resulting problem, we develop a penalty-based difference-of-convex successive convex approximation (PBDC-SCA) algorithm. Simulation results demonstrate that the proposed dynamic SC-MiLAC achieves task completion latency close to that of the fully digital architecture while outperforming static SC-MiLAC and conventional analog beamforming schemes.
	\end{abstract}
	

	\section{Introduction}
	Mobile edge computing (MEC) has emerged as a promising paradigm for supporting computation-intensive and latency-sensitive applications in wireless networks \cite{7879258,8016573,7488250}. By deploying computing resources at the network edge, MEC enables mobile devices to offload computational tasks to nearby edge servers, thereby reducing the computational burden on resource-constrained devices and improving task execution efficiency \cite{7572018,7762913}. However, mobile devices typically have limited battery capacity, and both local computing and wireless task offloading consume considerable energy, which may hinder the sustained operation of MEC services. By integrating wireless power transfer (WPT) with MEC, wireless powered MEC (WP-MEC) systems can alleviate the energy constraints of mobile devices and support sustained task execution. In a typical WP-MEC system, a base station (BS) first delivers radio-frequency (RF) energy to mobile users through downlink WPT, which is then used to support local computing and uplink task offloading \cite{8234686,8334188,8794550,7130662}. Nevertheless, the limited harvested energy and the coupled nature of energy harvesting (EH), task offloading, and computing resource allocation make efficient system design essential for reducing task completion latency.
	
	\subsection{Prior Works}
	In WP-MEC systems, the beamforming capability of the BS plays an important role in improving energy transfer efficiency. By exploiting multiple antennas, beamforming can spatially concentrate RF energy toward users, thereby increasing their received power and reducing the time required for EH \cite{6954434,6489506}. This is particularly important in multi-user scenarios, where users experience different channel conditions and require different amounts of energy to support their subsequent computing and offloading tasks \cite{9180255}. A fully digital (FD) architecture offers high beamforming flexibility by equipping each antenna with a dedicated radio frequency (RF) chain. However, the resulting hardware cost and power consumption can become substantial as the number of antennas increases \cite{7400949,8932399}. In contrast, analog beamforming (ABF) architectures reduce the number of RF chains by employing phase-shifter networks, but their beamforming flexibility is constrained by the limited control capabilities of analog phase shifters \cite{7400949,8932399}. In particular, a fixed analog configuration may not effectively accommodate the diverse channel conditions of multiple users or the different beamforming requirements of the WPT and wireless information transmission (WIT) phases. Therefore, developing a hardware-efficient architecture that provides sufficient beamforming flexibility while reducing RF-chain overhead is of significant interest for WP-MEC systems.
	
	To overcome the limitations of conventional beamforming architectures, microwave linear analog computers (MiLACs) have recently emerged as a promising solution for analog signal processing in wireless communications \cite{11281875,11281878,11431737,11455580}. Unlike conventional phase-shifter-based networks, a MiLAC performs linear transformations through a reconfigurable multiport microwave network with tunable susceptances, enabling flexible analog beamforming with a limited number of RF chains \cite{11281875,11281878,11431737}. Two representative MiLAC topologies are the fully connected (FC) and stem-connected (SC) architectures \cite{zhang2026beamforming,zhang2026many}. In FC-MiLAC, all RF-chain and antenna ports are interconnected through tunable circuit elements, providing flexible signal processing capabilities but resulting in hardware complexity that scales quadratically with the number of antennas. In contrast, SC-MiLAC employs a smaller fully connected core to which the remaining antenna ports are connected, reducing the number of tunable components to linear scaling with the antenna count for a fixed number of RF chains, while retaining key beamforming and capacity properties under ideal hardware assumptions \cite{11455580}. In particular, an ideal single-RF-chain SC-MiLAC can realize arbitrary unit-norm beamforming vectors, enabling flexible transmit and receive beamforming with substantially reduced RF-chain requirements \cite{11455580}. This capability is particularly attractive for WP-MEC systems, where downlink WPT must accommodate diverse user energy demands, while uplink task offloading requires user-specific receive combining. By dynamically reconfiguring its analog network, SC-MiLAC provides a hardware-efficient solution for supporting both energy transmission and uplink reception in WP-MEC systems.
	
	\subsection{Motivations}
	Despite the recent advances in WP-MEC and MiLAC technologies, their integration and joint optimization remain largely unexplored. Existing WP-MEC studies have investigated various techniques to improve energy transfer efficiency and reduce task completion latency, including nonlinear EH models and joint optimization of energy transmission, task offloading, and computing resources \cite{10107791,park2021latency}. Meanwhile, time-division energy beamforming has been studied to exploit the nonlinear RF-to-DC conversion characteristics in multiuser WPT systems \cite{9180255}. However, these studies do not account for the hardware characteristics and configuration-switching overhead of MiLAC-based analog beamforming. On the other hand, existing MiLAC studies have mainly focused on multiple-input multiple-output (MIMO) capacity analysis and downlink beamforming design \cite{11281878,zhang2026beamforming,11455580}, while its application to WP-MEC systems involving EH and computation offloading has received little attention. In particular, several challenges arise when employing SC-MiLAC in WP-MEC systems. \textit{First}, dynamically reconfiguring the MiLAC during WPT can provide additional flexibility in energy beamforming, but each configuration change incurs a switching delay during which energy transmission is suspended. The resulting tradeoff between energy beamforming gains and switching overhead needs to be carefully characterized and optimized, particularly under nonlinear EH. \textit{Second}, the MiLAC must also be reconfigured for uplink reception under the adopted time-division duplexing (TDD) protocol, introducing additional switching overhead across users' offloading slots. \textit{Third}, the harvested energy couples WPT beamforming and time allocation with local computing, task offloading, and edge computing resource allocation. The resulting optimization problem is further complicated by the nonlinear EH model and binary offloading decisions. Therefore, it is essential to develop an effective joint optimization framework that accounts for both the analog hardware constraints and switching overhead, while coordinating energy transfer, communication, and computation to minimize the maximum task completion latency.
	
	\subsection{Contributions}
	To address the aforementioned challenges, this paper investigates the integration of a hardware-efficient SC-MiLAC into WP-MEC systems to improve energy transfer efficiency and reduce task completion latency. The main contributions are summarized as follows.
	\begin{itemize}
		\item[$\bullet$] We establish a SC-MiLAC-enabled WP-MEC system that incorporates dynamic WPT beamforming, TDMA-based uplink reception, and hardware switching delays. Based on this model, we formulate a joint optimization problem to minimize the maximum task completion latency by optimizing WPT beamforming and time allocation, uplink transmission time and power, task offloading decisions, and local and edge computing resources, subject to nonlinear EH and SC-MiLAC hardware constraints.
		\item[$\bullet$] We provide a fundamental performance analysis of the MiLAC-enabled WP-MEC network, which reveals that: 1) FC- and SC-MiLAC achieve identical WPT performance under static configurations and ideal hardware assumptions; 2) under linear EH and negligible switching delays, a single-RF-chain SC-MiLAC achieves the same WPT performance and minimum task completion latency as fully digital architectures using no more beam configurations than users; 3) under nonlinear EH and finite switching delays, optimal dynamic WPT can be achieved with at most one more beam configuration than the number of users, and we derive a sufficient condition for its harvested energy to be no less than that of static fully digital WPT with the same average transmit covariance; and 4) under linear EH, we establish an upper bound on the additional task completion latency caused by hardware switching delays.
		\item[$\bullet$] To solve the formulated latency minimization problem, we develop a penalty-based difference-of-convex successive convex approximation (PBDC-SCA) algorithm. An alternating optimization framework is adopted to alternately optimize the WPT beamforming variables and the joint time and resource allocation variables. Specifically, DC-SCA is employed to construct tractable convex approximations for the nonconvexities associated with the inverse nonlinear EH model, received RF power constraints, and edge computing constraints. Meanwhile, a PB method is introduced to handle binary task offloading decisions by relaxing the binary constraints and penalizing nonbinary solutions.
		\item[$\bullet$] Numerical results demonstrate that: 1) ideal dynamic single-RF-chain SC-MiLAC achieves a task completion latency close to that of the fully digital scheme, while a practical switching delay of \(0.1~\rm ms\) causes only a slight latency increase, with practical dynamic SC-MiLAC still outperforming static SC-MiLAC, dynamic ABF, and conventional ABF; 2) the performance of dynamic SC-MiLAC depends on the number of WPT beam configurations and hardware switching delay, reflecting a tradeoff between beamforming flexibility and switching overhead; and 3) the performance gap between dynamic and fixed-configuration schemes widens with increasing numbers of users and task sizes, whereas higher BS transmit power reduces the performance gaps among different architectures.
	\end{itemize}
	
	\subsection{Organization and Notation}
	The remainder of this paper is organized as follows. Section II presents the SC-MiLAC-enabled WP-MEC system model. Section III analyzes the WPT and uplink reception capabilities of MiLAC architectures, including the impact of hardware switching delays on EH and task completion latency. Section IV formulates the latency minimization problem and develops the PBDC-SCA algorithm with complexity analysis. Section V provides numerical results to evaluate the proposed scheme, and Section VI concludes the paper.
	
	\textit{Notations:} Lowercase and uppercase boldface letters denote vectors and matrices, respectively, while italic letters denote scalars. $\mathbb{C}^{m\times n}$ and $\mathbb{R}^{m\times n}$ represent complex and real matrix spaces, respectively, and $\mathbf{I}_n$ is the identity matrix. The operators $(\cdot)^T$, $(\cdot)^H$, and $(\cdot)^*$ denote transpose, Hermitian transpose, and complex conjugate, respectively. $\operatorname{Tr}(\cdot)$, $\operatorname{rank}(\cdot)$, and $\operatorname{diag}(\cdot)$ denote trace, rank, and diagonal matrix formation. Moreover, $\|\cdot\|_2$, $\mathbb{E}[\cdot]$, $\mathcal{CN}(\boldsymbol{\mu},\mathbf{\Sigma})$, and $\operatorname{Re}\{\cdot\}$ denote the 2-norm, expectation, complex Gaussian distribution, and real part, respectively. The indicator $1_{\{\mathcal{A}\}}$ equals one if $\mathcal{A}$ holds and zero otherwise.
	
	\section{System Model}
	\begin{figure}[!t]
		\centering
		\includegraphics[width=3.2in]{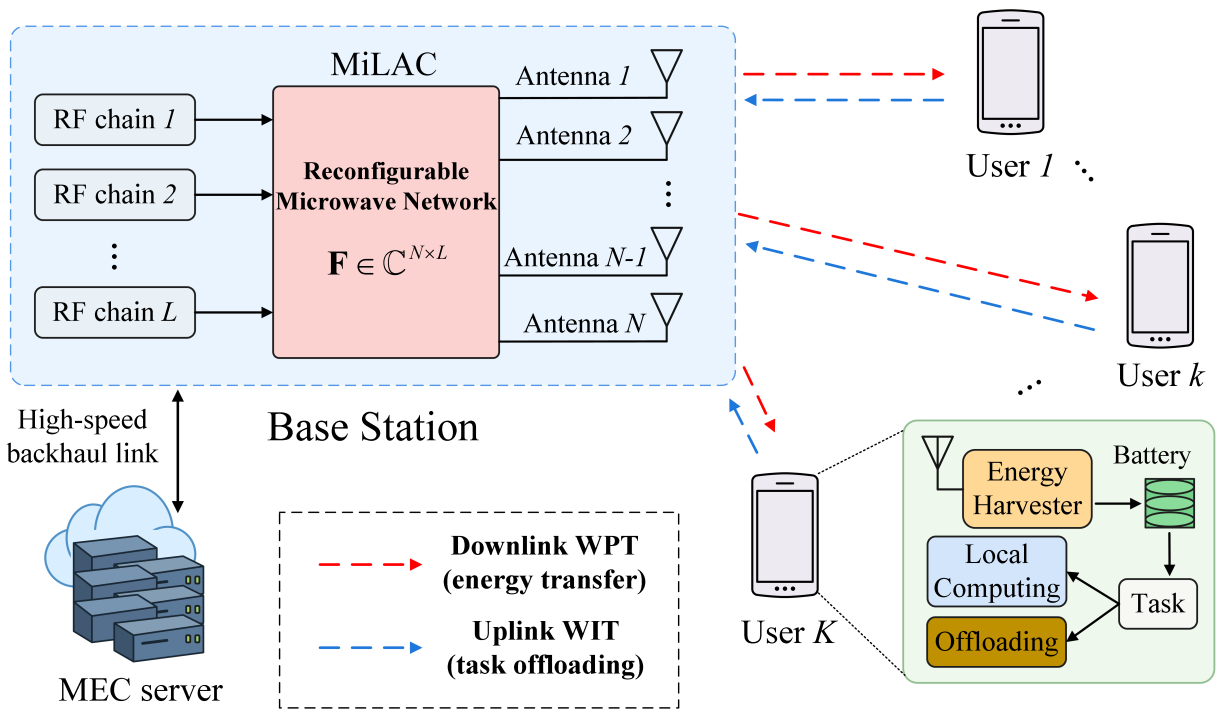}
		\caption{System architecture of the MiLAC-enabled WP-MEC system.}
		\label{system_model}
	\end{figure}
	As illustrated in Fig. 1, we consider a MiLAC-enabled WP-MEC system consisting of a BS equipped with \(N\) antennas and \(L\) RF chains, \(K\) single-antenna users, and an MEC server connected to the BS via a high-speed backhaul link. A MiLAC is deployed between the RF chains and the antenna array to perform analog beamforming. Each user has a computation task that is partitioned into a locally executed part and an offloaded part. The wireless transmission protocol consists of a downlink WPT phase of duration \(t_0\), followed by uplink TDMA offloading slots. The WPT phase is dedicated to EH. After the WPT phase, local computing starts at the beginning of the WIT phase and proceeds in parallel with uplink TDMA offloading. The MEC server processes each offloaded task upon reception, and local and edge computing may continue beyond the wireless transmission phase until all tasks are completed.
	
	A MiLAC is a reconfigurable microwave network that implements analog signal processing through wave propagation and tunable circuit elements \cite{11281878}. For the considered system, the MiLAC has \(N+L\) ports, comprising \(L\) RF-chain ports and \(N\) antenna ports. Let \(\boldsymbol{\Theta}\in\mathbb{C}^{(N+L)\times(N+L)}\) and \(\mathbf B\in\mathbb{R}^{(N+L)\times(N+L)}\) denote the scattering matrix and the real symmetric susceptance matrix of the MiLAC, respectively. Under the lossless and reciprocal assumptions, the scattering matrix is given by \(\boldsymbol{\Theta}=(Y_0\mathbf I-j\mathbf B)(Y_0\mathbf I+j\mathbf B)^{-1}\), where \(Y_0\) is the reference admittance. By ordering the RF-chain ports before the antenna ports, the effective analog beamforming matrix is defined as \(\mathbf F=[\boldsymbol{\Theta}]_{L+1:L+N,\,1:L}\in\mathbb C^{N\times L}\), which satisfies \(\|\mathbf F\|_2\leq 1\). We adopt a power-normalized traveling-wave representation throughout this work. The same reciprocal network supports downlink energy transmission and uplink signal reception \cite{zhang2026beamforming}.
	
	\begin{figure}[!t]
		\centering
		\includegraphics[width=3.2in]{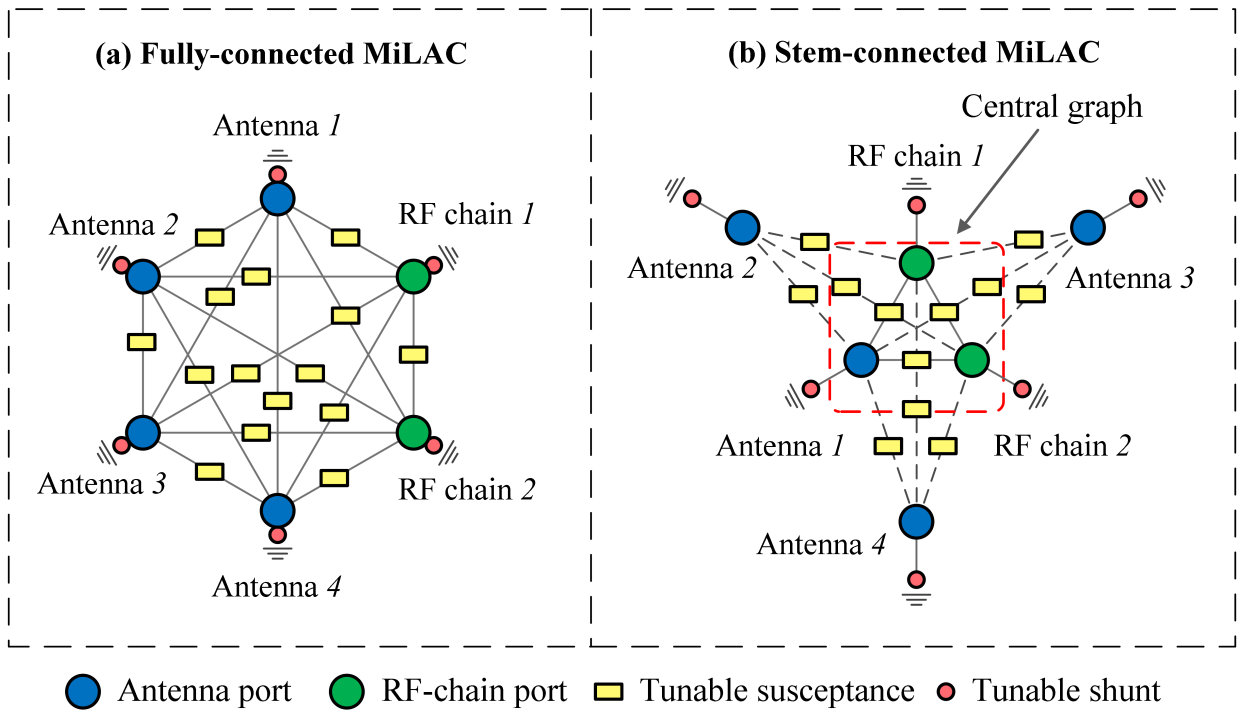}
		\caption{Topologies of (a) fully-connected and (b) stem-connected MiLAC architectures.}
		\label{MiLAC_topologies}
	\end{figure}
	
	Fig. \ref{MiLAC_topologies} compares the FC- and SC-MiLAC architectures. In the FC architecture, every pair of ports is interconnected through a tunable susceptance, with an additional tunable shunt element at each port. Its total number of tunable components is \((N+L)(N+L+1)/2\). In contrast, the SC architecture forms a central graph comprising all \(L\) RF-chain ports and \(L-1\) antenna ports, while each of the remaining \(N-L+1\) antenna ports is connected only to the central graph. This structure reduces the number of tunable components to \(L(2N+1)\). For a fixed number of RF chains, the SC architecture reduces the number of tunable components from quadratic to linear scaling with \(N\) \cite{zhang2026many}. Therefore, we adopt the SC-MiLAC to reduce the hardware complexity of the BS.

	\subsection{Proposed SC-MiLAC-Aided WPT}
	\begin{figure*}[!t]
		\centering
		\includegraphics[width=7.2in]{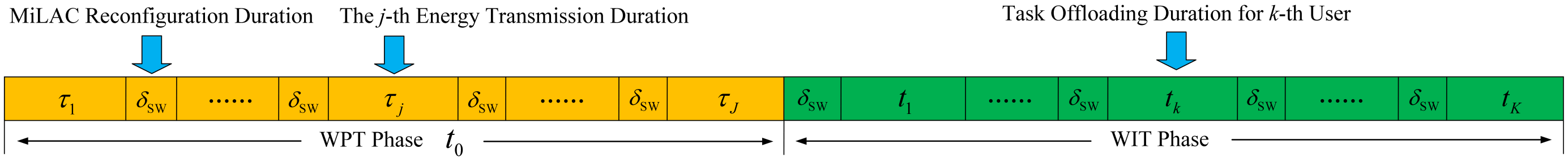}
		\caption{Illustration of the time allocation and beam switching during the WPT and WIT phases.}
		\label{WPT}
	\end{figure*}
	In multiuser WPT systems, maintaining a fixed energy beam throughout the WPT phase may limit the flexibility of energy allocation among users, particularly under nonlinear EH models. Previous studies have demonstrated that time-division energy beamforming can improve the harvested DC energy by exploiting the nonlinear RF-to-DC conversion characteristics \cite{9180255}. Motivated by this, we adopt a dynamic beam-switching scheme, where the SC-MiLAC adjusts its tunable susceptances to realize different energy beamforming configurations \cite{11455580}. Recent experimental studies on field-programmable gate array (FPGA)-controlled reconfigurable microwave hardware have demonstrated configuration update times on the order of \(0.1~{\rm ms}\) \cite{akram2026scalable}. As shown in Fig. \ref{WPT}, the WPT phase of duration \(t_0\) consists of \(J\) energy transmission subintervals, with \(\tau_j\) denoting the duration of the \(j\)-th subinterval. Accounting for the MiLAC reconfiguration time \(\delta_{\mathrm{sw}}\) between consecutive subintervals, we have \(\sum_{j=1}^{J}\tau_j+(J-1)\delta_{\mathrm{sw}} = t_0\). During each switching interval, RF energy transmission is suspended. The initial MiLAC configuration is assumed to be prepared before the WPT phase.
	During each subinterval, the SC-MiLAC realizes an analog beamforming matrix \(\mathbf F_j\in\mathbb C^{N\times L}\). 
	Let $\mathbf s_j\in\mathbb C^L$ denote the zero-mean RF-chain input signal. We assume that the input energy signals are mutually uncorrelated, with covariance matrix $\mathbf Q_j=\mathbb E[\mathbf s_j\mathbf s_j^H]=\operatorname{diag}(q_{j,1},\ldots,q_{j,L})$, where $q_{j,l}\ge0$ denotes the input power of RF chain $l$.
	The transmitted signal is given by \(\mathbf x_j=\mathbf F_j\mathbf s_j\), subject to the maximum transmit-power constraint
	\begin{flalign}
		\operatorname{Tr}
		\left(\mathbf F_j\mathbf Q_j\mathbf F_j^H\right)
		\le P_{\max},\quad \forall j, 
	\end{flalign}
	where \(P_{\max}\) denotes the maximum transmit power of the BS. Here, \(\mathbf F_j\) must be realizable by the SC-MiLAC configuration. Let \(\mathbf p_{\mathrm{BS}}=[0,0,h_{\mathrm{BS}}]^T\) and \(\mathbf p_k=[x_k,y_k,0]^T\) denote the positions of the BS array center and user \(k\), respectively, with \(d_k=\|\mathbf p_k-\mathbf p_{\mathrm{BS}}\|_2\). We adopt a Rician fading channel with large-scale path loss, expressed as
	\begin{flalign}
		{{\bf{h}}_k} = \sqrt {{\beta _k}} \left( {\sqrt {\frac{{{\kappa _k}}}{{{\kappa _k} + 1}}} {\bf{a}}\left( {{\mathbf p_k}} \right) + \sqrt {\frac{1}{{{\kappa _k} + 1}}} {{\bf{\bar h}}_k}} \right),
	\end{flalign}
	where \(\beta_k=\beta_0(d_k/d_0)^{-\alpha}\) denotes the large-scale channel power gain, with \(\beta_0\) being the reference channel power gain at distance \(d_0\) and \(\alpha\) the path-loss exponent. The parameter \(\kappa_k\) is the Rician factor in linear scale, while \(\mathbf {\bar h}_k\sim\mathcal{CN}(\mathbf 0,\mathbf I_N)\) represents the non-line-of-sight (NLoS) component. For a horizontal uniform linear array (ULA) with inter-antenna spacing \(d_{\mathrm a}=\lambda/2\), the position of the \(n\)-th antenna is \(\mathbf p_{\mathrm{BS},n}=\mathbf p_{\mathrm{BS}}+[(n-(N+1)/2)d_{\mathrm a},0,0]^T\). The spherical-wave line-of-sight (LoS) array response is given by \([\mathbf a(\mathbf p_k)]_n=\frac{d_k}{d_{k,n}}e^{-j\frac{2\pi}{\lambda}(d_{k,n}-d_k)}\), where \(d_{k,n}=\|\mathbf p_k-\mathbf p_{\mathrm{BS},n}\|_2\), \(n=1,\ldots,N\), and \(\lambda\) denotes the carrier wavelength. The channel is assumed to remain constant within one frame.
	
	During the \(j\)-th WPT subinterval, the received signal at user \(k\) is \(y_{k,j}^{\mathrm{WPT}}=\mathbf h_k^H\mathbf F_j\mathbf s_j+n_{k,j}\), where \(n_{k,j}\) denotes the receiver noise. Neglecting the noise contribution to EH, the average received RF power is given by
	\begin{flalign}
		P_{k,j}^{{\rm{RF}}} = {\mathbb E}\left[ {{{\left| {{\bf{h}}_k^H{{\bf{F}}_j}{{\bf{s}}_j}} \right|}^2}} \right] = {\bf{h}}_k^H{{\bf{F}}_j}{{\bf{Q}}_j}{\bf{F}}_j^H{{\bf{h}}_k}.
	\end{flalign}
	
	To account for the nonlinear characteristics of practical EH circuits, we adopt the normalized sigmoid model \cite{park2021latency,9957071,11206485}. Specifically, the harvested DC output power of user \(k\) is given by
	\begin{flalign}
		P_{k,j}^{{\rm{DC}}} = \frac{{{M_k}}}{{1 - {\Omega _k}}}\left[ {\frac{1}{{1 + \exp \left[ { - {a_k}\left( {P_{k,j}^{{\rm{RF}}} - {b_k}} \right)} \right]}} - {\Omega _k}} \right],
	\end{flalign}
	where \(M_k\) denotes the saturation DC output power, \(a_k\) and \(b_k\) are circuit-dependent parameters, and \(\Omega_k=[1+\exp(a_kb_k)]^{-1}\) ensures that the harvested DC output power is zero when the RF input power is zero. Accordingly, the total energy harvested by user \(k\) over the WPT phase is \(E_k^{{\rm{EH}}} = \sum\nolimits_{j = 1}^J {{\tau _j}P_{k,j}^{{\rm{DC}}}} \). The harvested energy is stored in the users' batteries and used to support local computing and uplink task offloading during the subsequent WIT phase.

	\subsection{Computation Offloading and Task Execution}
	We consider a partial computation offloading scheme, where each user \(k\) has a computation task of \(D_k\) bits, with \(c_k\) CPU cycles required per bit. Let \(\ell_k\in[0,D_k]\) denote the number of bits offloaded to the MEC server, while the remaining \(D_k-\ell_k\) bits are processed locally. During the uplink WIT phase, the users sequentially offload their tasks through TDMA, where \(t_k\) and \(p_k\) denote the transmission duration and power of user \(k\), respectively.
	
	\subsubsection{Uplink computation offloading} 
	Assuming calibrated time-division duplexing (TDD) channel reciprocity, the uplink channel is given by \(\mathbf g_k=\mathbf h_k^*\in\mathbb C^{N\times1}\). During user \(k\)'s offloading slot, the SC-MiLAC is reconfigured to realize the analog beamforming matrix \(\mathbf F_k^{\mathrm{UL}}\in\mathbb C^{N\times L}\), whose transpose represents the corresponding uplink analog combining matrix due to reciprocity. Let \(\mathbf w_k\in\mathbb C^L\) denote the digital combining vector applied to the RF-chain outputs, yielding the effective receive combining vector \(\mathbf v_k=(\mathbf F_k^{\mathrm{UL}})^*\mathbf w_k\). Accordingly, the achievable uplink transmission rate is given by
	\begin{flalign}
		R_k^{{\rm{UL}}} = W{\log _2}\left( {1 + \frac{{{p_k}{{\left| {{\bf v}_k^H{\bf g}_k} \right|}^2}}}{{\sigma _{{\rm{BS}}}^2{{\left\| {{{\bf v}_k}} \right\|}^2}}}} \right),
	\end{flalign}
	where \(W\) denotes the transmission bandwidth and \(\sigma_{\mathrm{BS}}^2\) is the noise power per antenna. Thus, the offloaded task size satisfies \(\ell_k\le t_kR_k^{\mathrm{UL}}\). Accounting for the fixed circuit power \(P_{k,\mathrm{cir}}^{\mathrm{UL}}\), the energy consumed during task offloading is \(E_k^{\mathrm{UL}}=(p_k+P_{k,\mathrm{cir}}^{\mathrm{UL}})t_k\). The transmit power of user \(k\) is constrained by \(0\le p_k\le p_{k,\max}\), where \(p_{k,\max}\) denotes its maximum transmit power.
	
	\subsubsection{Local computing}
	Each user employs dynamic voltage and frequency scaling (DVFS) for local computing, with CPU frequency \(f_k^{\mathrm{loc}}\le f_{k,\max}^{\mathrm{loc}}\). The CPU frequency is assumed constant during the execution of each local computing task. Since the harvested energy is stored in the users' batteries, local computing starts at the beginning of the WIT phase and proceeds in parallel with uplink offloading \cite{11618930}. The local computation time and energy consumption are respectively given by
	\begin{flalign}
		t_k^{{\rm{loc}}} &= \frac{{{c_k}\left( {{D_k} - {\ell_k}} \right)}}{{f_k^{\rm loc}}},\\
		E_k^{\rm loc} &= \left[\xi _k \left(f_k^{\rm loc}\right)^3 + P_{k,{\rm cir}}^{\rm CPU}\right] t_k^{\rm loc},
	\end{flalign}
	where \(\xi_k\) denotes the effective switched-capacitance coefficient and \(P_{k,\mathrm{cir}}^{\mathrm{CPU}}\) represents the fixed circuit power consumed during local computing.
	
	\subsubsection{Edge computing and task completion}
	After receiving the offloaded task from user \(k\), the MEC server allocates a CPU frequency \(f_k^{\mathrm{edge}}\) to execute it, subject to \(\sum_{k=1}^{K}f_k^{\mathrm{edge}}\le F_{\max}\), where \(F_{\max}\) denotes the total computing capacity of the MEC server. We assume that different offloaded tasks can be processed concurrently using their allocated constant CPU resources. Let \(u_k\) denote the completion time of user \(k\)'s uplink transmission. 
	Accounting for a MiLAC reconfiguration delay $\delta_{\mathrm{sw}}$ before each active uplink offloading slot, the completion time of user $k$'s uplink transmission is given by \(u_k=t_0+\sum_{i=1}^{k}
	\left(t_i+\delta_{\mathrm{sw}}\mathbf{1}_{\{\ell_i>0\}}\right), \ell_k>0\),
	where $\mathbf{1}_{\{\ell_i>0\}}$ is an indicator function that equals one if $\ell_i>0$ and zero otherwise, with $t_i=0$ when $\ell_i=0$. The corresponding edge computation time is $t_k^{\mathrm{edge}}=c_k\ell_k/f_k^{\mathrm{edge}}$ for $\ell_k>0$.
	Neglecting the backhaul transmission and result downloading delays, the task completion time of user \(k\) is given by
		\begin{flalign}
			T_k &= \max\left\{t_0+t_k^{\mathrm{loc}},
			T_k^{\mathrm{off}}\right\}, \\
			T_k^{\mathrm{off}}
			&=
			\begin{cases}
				u_k+\dfrac{c_k\ell_k}{f_k^{\mathrm{edge}}},
				& \ell_k>0,\\
				0, & \ell_k=0.
			\end{cases}
		\end{flalign}
	Here, \(t_k^{\mathrm{loc}}=0\) when \(\ell_k=D_k\), and the CPU frequencies corresponding to nonzero workloads are assumed positive.
	
	Since WPT precedes the WIT phase, the harvested energy is available in the users' batteries at the beginning of the WIT phase. We assume a sufficiently large battery capacity to avoid energy overflow. Following a frame-level energy-neutral operation, the energy consumed for local computing and task offloading is constrained by
	\begin{flalign}
		E_k^{\rm loc} + E_k^{\rm UL} \le E_k^{\rm EH}, \forall k.
	\end{flalign}

	\section{Performance Analysis for SC-MiLAC Enabled WP-MEC networks}
	
	In this section, we first characterize the WPT covariance realizability and RF-chain requirements of FC- and SC-MiLAC architectures. We then analyze single-RF-chain dynamic WPT under linear and nonlinear EH with switching delays, and investigate uplink reception and the resulting task completion latency.
	
	\subsection{WPT Beamforming Capability and RF-Chain Requirements}
	
	For static WPT, we omit the subinterval index \(j\) and express the transmit covariance matrix as \(\mathbf S=\mathbf F\mathbf Q\mathbf F^H\), where \(\mathbf Q\) is diagonal and \(\operatorname{Tr}(\mathbf S)\le P_{\max}\). Let $\mathcal S_L^{\mathrm{FC}}$ and $\mathcal S_L^{\mathrm{SC}}$ denote the sets of WPT transmit covariance matrices achievable by FC- and SC-MiLAC architectures with $L$ RF chains, respectively.
	An ideal SC-MiLAC can realize any semi-unitary matrix \(\mathbf F\in\mathbb C^{N\times L}\) satisfying \(\mathbf F^H\mathbf F=\mathbf I_L\) \cite{11455580,zhang2026beamforming}. Since the FC architecture contains the SC topology as a special case, the achievable covariance sets of both architectures are identical, i.e.,
	\begin{equation}
		\label{eq:covariance}
		\begin{aligned}
			\!\!\mathcal S_L^{\mathrm{FC}}=\mathcal S_L^{\mathrm{SC}}=\Big\{\mathbf S\succeq\mathbf0:
			\operatorname{rank}(\mathbf S)\le L,\operatorname{Tr}(\mathbf S)\le P_{\max}\Big\}.
		\end{aligned}
	\end{equation}
	
	Consequently, \textit{SC- and FC-MiLAC architectures achieve identical static WPT performance}.
	
	\subsection{Single-RF-Chain Dynamic WPT}
	From \eqref{eq:covariance}, a nonzero covariance matrix \(\mathbf S\) requires at least \(L_{\min}=\operatorname{rank}(\mathbf S)\) RF chains. Let $\mathbf S_{\mathrm{FD}}^\star$ denote a minimum-rank optimal covariance matrix of a fully digital static WPT design. Since projecting any feasible covariance onto the column space of $\mathbf H=[\mathbf h_1,\ldots,\mathbf h_K]$ preserves every user's received RF power without increasing the transmit power, the minimum RF-chain requirement for achieving the fully digital static WPT performance satisfies
	\begin{equation}
		\label{eq:min_rf}
		L_{\min}^{\mathrm{stat}}
		=\operatorname{rank}(\mathbf S_{\mathrm{FD}}^\star)
		\le\operatorname{rank}(\mathbf H)
		\le\min\{K,N\}.
	\end{equation}
	
	The preceding analysis shows that static WPT may require multiple RF chains to realize a high-rank transmit covariance matrix. Can dynamic beam switching eliminate the need for multiple RF chains? To address this question, we first consider an ideal linear EH model with negligible switching delay, i.e., $\delta_{\mathrm{sw}}=0$. Under this model, the harvested energy of user $k$ is $E_k^{\mathrm{EH}}=\eta_k t_0\mathbf h_k^H\overline{\mathbf S}\mathbf h_k$, where $\eta_k\in(0,1]$ denotes the RF-to-DC conversion efficiency and $\overline{\mathbf S}$ is the time-averaged transmit covariance matrix.
	
	\begin{theorem}
		For an ideal single-RF-chain SC-MiLAC, the following properties hold: (a) Under linear EH and negligible switching delays, dynamic WPT achieves the same per-user harvested energy as fully digital WPT using at most $K$ beam configurations. (b) Even with nonlinear EH and finite switching delays, optimal dynamic WPT performance can be achieved using at most $K+1$ beam configurations.
		\label{thm:dynamic_wpt}
	\end{theorem}
	
	\begin{proof}
		For Theorem 1 (a), consider any feasible fully digital WPT scheme with time-averaged covariance $\overline{\mathbf S}=\sum_{j=1}^{r}\lambda_j\mathbf u_j\mathbf u_j^H$, where $r=\operatorname{rank}(\overline{\mathbf S})$, $\lambda_j>0$, and $\|\mathbf u_j\|=1$. For $P=\operatorname{Tr}(\overline{\mathbf S})>0$, construct a single-RF-chain scheme with
		\begin{equation}
			\label{eq:dynamic_beam}
			\mathbf f_j=\mathbf u_j,\quad
			q_j=P,\quad
			\tau_j=\frac{t_0\lambda_j}{P},
			\quad j=1,\ldots,r.
		\end{equation}
		Since $P\le P_{\max}$ and $\sum_{j=1}^{r}\tau_j=t_0$, the constructed scheme satisfies the power and time constraints. Moreover,
		\begin{equation}
			\label{eq:average_covariance}
			\frac{1}{t_0}\sum\nolimits_{j=1}^{r}
			\tau_jq_j\mathbf f_j\mathbf f_j^H
			=\sum\nolimits_{j=1}^{r}
			\lambda_j\mathbf u_j\mathbf u_j^H
			=\overline{\mathbf S}.
		\end{equation}
		
		Each unit-norm vector $\mathbf f_j$ is realizable by an ideal single-RF-chain SC-MiLAC \cite{11455580}. Thus, every user obtains the same received energy as in the fully digital scheme. Furthermore, projecting $\overline{\mathbf S}$ onto the column space of $\mathbf H$ preserves every user's received RF power without increasing the transmit power. Applying the same construction to the projected covariance requires at most $\operatorname{rank}(\mathbf H)\le K$ beam configurations. The zero-covariance case is trivially achievable by remaining idle. Since any single-RF-chain scheme is also feasible for a fully digital transmitter, the two architectures achieve the same WPT performance.
		
		For Theorem 1(b), consider any feasible single-RF-chain dynamic WPT scheme with $J$ beam configurations. Let $\mathbf e_j=[P_{1,j}^{\mathrm{DC}},\ldots,P_{K,j}^{\mathrm{DC}}]^T$ denote the harvested DC power vector under configuration $j$. The total harvested energy satisfies
		\begin{equation}
			\label{eq:energy_convex}
			\frac{1}{t_{\mathrm{tx}}}\mathbf E^{\mathrm{EH}}
			=\sum\nolimits_{j=1}^{J}\frac{\tau_j}{t_{\mathrm{tx}}}\mathbf e_j,
		\end{equation}
		where $t_{\mathrm{tx}}=\sum_{j=1}^{J}\tau_j>0$ and $\mathbf E^{\mathrm{EH}}=[E_1^{\mathrm{EH}},\ldots,E_K^{\mathrm{EH}}]^T$. Since the right-hand side of \eqref{eq:energy_convex} is a convex combination of vectors in $\mathbb R^K$, Carath\'eodory's theorem guarantees that the same harvested energy can be achieved using at most $K+1$ of the original beam configurations with appropriately adjusted durations. This construction preserves the total energy transmission time while reducing or maintaining the number of switching intervals. Consequently, the required WPT duration does not increase, and an optimal solution with at most $K+1$ beam configurations always exists.
	\end{proof}
	
	\subsection{Nonlinear EH and Switching Overhead}
	Under the adopted sigmoid EH model, \(\Phi_k(P)\) is convex for \(P<b_k\) and concave for \(P>b_k\). Thus, variations in received RF power can increase the average DC output when all received powers lie in the convex region, but reduce it when they lie in the concave region. In either case, the switching overhead must also be considered.
	
	To analyze this tradeoff, consider $J\ge2$ beam configurations over a WPT duration $t_0$. Accounting for the switching delay, the effective energy transmission time is $t_{\mathrm{tx}}=t_0-(J-1)\delta_{\mathrm{sw}}>0$, assuming that the available transmission time is fully utilized. Let $\alpha_j=\tau_j/t_{\mathrm{tx}}$ denote the active-time fraction of configuration $j$, with $\sum_{j=1}^{J}\alpha_j=1$. The average received RF power and harvested DC power of user $k$ are given by
	
	\begin{flalign}
		\bar P_k^{\mathrm{RF}}
		&=\sum\nolimits_{j=1}^{J}\alpha_j P_{k,j}^{\mathrm{RF}},\\
		\bar P_k^{\mathrm{DC}}
		&=\sum\nolimits_{j=1}^{J}\alpha_j\Phi_k(P_{k,j}^{\mathrm{RF}}). \label{eq14}
	\end{flalign}
	
	For comparison, consider a static fully digital WPT scheme that transmits with covariance \(\overline{\mathbf S}=\sum_{j=1}^{J}\alpha_j\mathbf S_j\) throughout the WPT duration \(t_0\), where \(\overline{\mathbf S}\) is the average transmit covariance of the dynamic scheme over its active transmission intervals. The harvested energy difference between the dynamic and static schemes is given by
	\begin{flalign}
			\Delta E_k^{\mathrm{EH}}
			&=E_{k,\mathrm{dyn}}^{\mathrm{EH}}
			-E_{k,\mathrm{FD}}^{\mathrm{EH}}\notag\\
			&=t_0\left[
			\bar P_k^{\mathrm{DC}}
			-\Phi_k(\bar P_k^{\mathrm{RF}})
			\right]-(J-1)\delta_{\mathrm{sw}}\bar P_k^{\mathrm{DC}}.  \label{eq15}
	\end{flalign}
	
	The first term represents the energy gain or loss caused by nonlinear EH, while the second term captures the switching loss. To ensure that dynamic WPT achieves no less harvested energy than the static fully digital scheme, the energy difference in \eqref{eq15} must satisfy \(\Delta E_k^{\rm EH}\geq 0\) for every user. For each user with \(\bar{P}_k^{\rm DC}>0\), substituting \eqref{eq15} and rearranging yields the following condition on the switching delay \(\delta_{\rm sw}\leq \delta_{k,\rm crit}=\frac{t_0}{J-1}\left(1-\frac{\Phi_k(\bar{P}_k^{\rm RF})}{\bar{P}_k^{\rm DC}}\right)\). Since this condition must hold for every user, the switching delay must not exceed the smallest user-specific threshold. Therefore, dynamic WPT achieves no less harvested energy for every user if and only if
	\begin{flalign}
		\delta_{\rm sw} \leq \delta_{\rm crit}
		= \frac{t_0}{J-1}
		\min_{k:\bar{P}_k^{\rm DC}>0}
		\left(
		1-\frac{{\Phi}_k\left(\bar{P}_k^{\rm RF}\right)}
		{\bar{P}_k^{\rm DC}}
		\right). \label{eq16}
	\end{flalign}
	
	Define the received RF power variance as \(V_k=\sum_{j=1}^{J}\alpha_j\left(P_{k,j}^{\mathrm{RF}}-\bar P_k^{\mathrm{RF}}\right)^2\) and the minimum curvature of the EH function over \(\mathcal I_k\) as \(m_k\triangleq\min_{P\in\mathcal I_k}\Phi_k''(P)\), where \(\mathcal I_k\) is the smallest closed interval containing all \(P_{k,j}^{\mathrm{RF}}\). Under the condition \(P_{k,j}^{\mathrm{RF}}<b_k\) for all \(k,j\), we have \(m_k>0\).
	
	\begin{theorem}
		Under nonlinear EH, dynamic WPT can achieve no less harvested energy for every user than its static fully digital counterpart with the same average transmit covariance, despite hardware switching delays. Specifically, if $P_{k,j}^{\rm RF}<b_k$, $\forall k,j$, this performance is guaranteed when
		\begin{flalign}
			\delta_{\rm sw} \leq \frac{t_0}{J-1}
			\min_{k:\bar{P}_k^{\rm DC}>0}
			\frac{m_k V_k}
			{2\Phi_k\left(\bar{P}_k^{\rm RF}\right)+m_k V_k}. \label{eq17}
		\end{flalign}
	\end{theorem}
	\begin{proof}
		Since \(P_{k,j}^{\rm RF}<b_k\) for all \(k,j\), the EH function is strongly convex over \(\mathcal I_k\), with \(\Phi_k''(P)\geq m_k>0\). Thus, the function \(\Phi_k(P)-\frac{m_k}{2}P^2\) is convex over \(\mathcal I_k\). Applying Jensen's inequality with \(\sum_{j=1}^{J}\alpha_j=1\) yields
		\begin{flalign}
			&\sum\nolimits_{j=1}^{J}\alpha_j\Phi_k(P_{k,j}^{\rm RF})
			-\Phi_k\left(\sum\nolimits_{j=1}^{J}\alpha_jP_{k,j}^{\rm RF}\right)\notag\\
			&\quad\geq \frac{m_k}{2}
			\left[
			\sum\nolimits_{j=1}^{J}\alpha_j(P_{k,j}^{\rm RF})^2
			-(\bar P_k^{\rm RF})^2
			\right]=\frac{m_kV_k}{2}, \label{eq18}
		\end{flalign}
		where the last equality follows from the definition of \(V_k\). Using \eqref{eq14}, we obtain
		\begin{flalign}
			\bar P_k^{\rm DC}
			-\Phi_k(\bar P_k^{\rm RF})
			\geq \frac{m_kV_k}{2}. \label{eq19}
		\end{flalign}
		Next, let \(s=(J-1)\delta_{\rm sw}\) denote the total switching time. From \eqref{eq15}, the harvested energy difference can be rewritten as
		\begin{flalign}
			\Delta E_k^{\rm EH}
			={}(t_0-s)
			\left[\bar P_k^{\rm DC}
			-\Phi_k(\bar P_k^{\rm RF})\right]-s\Phi_k(\bar P_k^{\rm RF}), \label{eq20}
		\end{flalign}
		where we have used \(\bar P_k^{\rm DC}=[\bar P_k^{\rm DC}-\Phi_k(\bar P_k^{\rm RF})]+\Phi_k(\bar P_k^{\rm RF})\).
		Since \(t_0-s>0\), applying \eqref{eq19} gives
		\begin{flalign}
			\Delta E_k^{\rm EH}
			\geq{}
			\frac{m_kV_k}{2}
			\left[t_0-(J-1)\delta_{\rm sw}\right]-(J-1)\delta_{\rm sw}
			\Phi_k(\bar P_k^{\rm RF}). \label{eq21}
		\end{flalign}
		
		For each user with \(\bar P_k^{\rm DC}>0\), the right-hand side of \eqref{eq21} is nonnegative if and only if
		\begin{flalign}
			\frac{m_kV_k}{2}
			\left[t_0-(J-1)\delta_{\rm sw}\right] \geq (J-1)\delta_{\rm sw}
			\Phi_k(\bar P_k^{\rm RF}), \label{eq22}
		\end{flalign}
		rearranging \eqref{eq22} yields
		\begin{flalign}
			\delta_{\rm sw}
			\leq \frac{t_0}{J-1}
			\frac{m_kV_k}
			{2\Phi_k(\bar P_k^{\rm RF})+m_kV_k}. \label{eq23}
		\end{flalign}
		
		Taking the minimum over all users with positive average harvested DC power gives \eqref{eq17}, which guarantees \(\Delta E_k^{\rm EH}\geq0\) for every such user. Users with \(\bar P_k^{\rm DC}=0\) harvest zero energy under both schemes. Therefore, the theorem follows.
	\end{proof}
	
	The user-specific switching-delay bound increases with \(m_kV_k\) for a fixed \(\Phi_k(\bar P_k^{\mathrm{RF}})\), while the system-wide bound is determined by the most restrictive user. In contrast, for any user whose received RF powers lie entirely in the concave region, Jensen's inequality gives \(\bar P_k^{\mathrm{DC}}\leq\Phi_k(\bar P_k^{\mathrm{RF}})\), and switching overhead further reduces its harvested energy.
	
	For given computation and offloading decisions, let \(E_k^{\mathrm{req}}=E_k^{\mathrm{loc}}+E_k^{\mathrm{UL}}\) denote the energy demand of user \(k\). With fixed beam configurations and active-time fractions, the minimum WPT duration required to satisfy all users' energy constraints is given by
	\begin{flalign}
		t_{0,\rm dyn}^{\min}
		= (J-1)\delta_{\rm sw}
		+ \max_{k:E_k^{\rm req}>0}
		\frac{E_k^{\rm req}}{\bar{P}_k^{\rm DC}}. \label{eq24}
	\end{flalign}
	For the corresponding static fully digital scheme with the same average transmit covariance, the required duration is
	\begin{flalign}
		t_{0,\mathrm{FD}}^{\min}
		=
		\max_{k:E_k^{\mathrm{req}}>0}
		\frac{E_k^{\mathrm{req}}}
		{\Phi_k(\bar P_k^{\mathrm{RF}})}. \label{eq25}
	\end{flalign}
	
	These expressions assume positive harvested DC power under both schemes for every user with a positive energy demand. Comparing \eqref{eq24} and \eqref{eq25}, dynamic WPT requires a shorter minimum WPT duration than the corresponding static fully digital scheme if and only if
	\begin{flalign}
		(J-1)\delta_{\rm sw}
		< t_{0,\rm FD}^{\min}
		- \max_{k:E_k^{\rm req}>0}
		\frac{E_k^{\rm req}}{\bar{P}_k^{\rm DC}}.
	\end{flalign}
	
	Since the required WPT duration depends on the computation and offloading decisions and precedes both local computing and uplink offloading, beam configurations, active-time fractions, and computation offloading decisions should be jointly optimized to minimize the overall task completion latency.
	
	\subsection{Uplink Reception and Task Completion Latency}
	We first examine whether reducing the number of RF chains affects uplink reception. Since TDMA is adopted, the BS receives signals from only one user at a time. For a single-RF-chain SC-MiLAC, let \(\mathbf f_k^{\mathrm{UL}}\in\mathbb C^N\) denote the analog combining vector for user \(k\), with \(\|\mathbf f_k^{\mathrm{UL}}\|=1\). The received signal-to-noise ratio (SNR) satisfies
	\begin{flalign}
		\gamma_k^{\mathrm{UL}}
		=
		\frac{p_k|(\mathbf f_k^{\mathrm{UL}})^T\mathbf g_k|^2}
		{\sigma_{\mathrm{BS}}^2}
		\le
		\frac{p_k\|\mathbf g_k\|^2}
		{\sigma_{\mathrm{BS}}^2},
	\end{flalign}
	where \(\mathbf g_k=\mathbf h_k^*\). The equality is achieved by choosing \(\mathbf f_k^{\mathrm{UL}}=\mathbf h_k/\|\mathbf h_k\|\), which is realizable by an ideal single-RF-chain SC-MiLAC. Thus, the maximum uplink rate is given by
	\begin{flalign}
		R_k^{\rm UL,\max} = W \log_2 \left(1+\frac{p_k \left\|{\bf h}_k\right\|^2}{\sigma_{{\rm{BS}}}^2}\right),
	\end{flalign}
	which is identical to that achieved by fully digital maximum-ratio combining (MRC). \textit{Therefore, a single RF chain does not reduce the achievable uplink rate under TDMA, provided that the analog combining vector can be reconfigured for each user}.
	
	Let \(T_{\mathrm{SC}}^\star(\delta_{\mathrm{sw}})\) and \(T_{\mathrm{FD}}^\star\) denote the minimum task completion latencies achieved by a single-RF-chain SC-MiLAC and a fully digital BS, respectively. The fully digital BS is assumed to incur no analog reconfiguration delay. 
	Consider an optimal fully digital solution with a nonzero WPT covariance \(\overline{\mathbf S}_{\mathrm{FD}}^\star\), chosen to have the minimum rank among optimal solutions. Define \(r^\star=\operatorname{rank}(\overline{\mathbf S}_{\mathrm{FD}}^\star)\) and \(K_{\mathrm{off}}^\star=|\{k:\ell_k^\star>0\}|\), where \(K_{\mathrm{off}}^\star\) is the number of offloading users.
	
	\begin{theorem}
		Under linear EH and ideal SC-MiLAC hardware assumptions, a single-RF-chain SC-MiLAC achieves the same minimum task completion latency as a fully digital architecture when switching delays are negligible. When switching delays are considered and the number of WPT beam configurations can be freely selected, the latency gap between the two architectures is upper-bounded by $(2K-1)\delta_{\mathrm{sw}}$.
	\end{theorem}
	\begin{proof}
		Consider the optimal fully digital solution. According to Theorem 1, its WPT covariance can be reproduced by a single-RF-chain SC-MiLAC using \(r^\star\) beam configurations under linear EH, while preserving every user's harvested energy. Since the WPT duration is a decision variable, the same active transmission time can be maintained by extending the WPT phase by \((r^\star-1)\delta_{\mathrm{sw}}\) to accommodate the switching intervals.
		
		For uplink offloading, the single-RF-chain SC-MiLAC achieves the same receive SNR as fully digital MRC by configuring its analog combining vector for each user. Hence, the same offloading decisions, transmit powers, and transmission durations remain feasible. According to the uplink timing model in Section II-B, serving the \(K_{\mathrm{off}}^\star\) offloading users introduces an additional delay of at most \(K_{\mathrm{off}}^\star\delta_{\mathrm{sw}}\).
		
		Under the adopted timing model, local computing starts at the end of the WPT phase. Thus, the additional WPT switching time delays both local computing and uplink offloading by \((r^\star-1)\delta_{\mathrm{sw}}\), while the uplink reconfiguration adds at most \(K_{\mathrm{off}}^\star\delta_{\mathrm{sw}}\) to an offloading path. Since the additional switching delays only postpone the start of local computing and the completion of uplink offloading, the original computing durations and resource allocations remain feasible. By keeping the local and edge computing allocations unchanged, the resulting task completion latency increases by at most \((r^\star-1+K_{\mathrm{off}}^\star)\delta_{\mathrm{sw}}\). Furthermore, \(r^\star\le\operatorname{rank}(\mathbf H)\le K\) and \(K_{\mathrm{off}}^\star\le K\). Since the fully digital BS can reproduce any feasible single-RF-chain transmission and reception strategy without analog reconfiguration delays, \(T_{\mathrm{FD}}^\star\le T_{\mathrm{SC}}^\star(\delta_{\mathrm{sw}})\). Combining these results completes the proof.
	\end{proof}
	
	In particular, setting \(\delta_{\mathrm{sw}}=0\) in Theorem 3 yields
	\begin{flalign}
		T_{\rm SC}^\star \left(0\right) = T_{\rm FD}^\star.
	\end{flalign}
	
	Thus, under linear EH and negligible switching delay, a single-RF-chain SC-MiLAC can achieve the same minimum task completion latency as a fully digital BS. Under nonlinear EH, however, the harvested energy also depends on the beam switching strategy, as analyzed in Section III-C.

	\section{Joint Beamforming and Resource Allocation for SC-MiLAC-Aided WP-MEC Systems}
	In this section, we develop an optimization framework to minimize the maximum task completion latency in the SC-MiLAC-aided WP-MEC system. The coupled beamforming, task offloading, and computing resource allocation variables are optimized through an alternating optimization approach. Specifically, we first optimize the WPT beamforming, followed by joint time and resource allocation with penalty-based offloading selection. Finally, the overall algorithm and its computational complexity are presented.
	
	\subsection{Problem Formulation and Reformulation}
	
	For a given number \(J\) of WPT beam configurations, let \(\mathbf f_j\in\mathbb C^{N\times1}\) and \(q_j\) denote the unit-norm transmit beamforming vector and transmit power of configuration \(j\), respectively. Since an ideal single-RF-chain SC-MiLAC can realize any unit-norm beamforming vector, the received RF power of user \(k\) under configuration \(j\) is \(P_{k,j}^{\mathrm{RF}}=q_j|\mathbf h_k^H\mathbf f_j|^2\), where \(\|\mathbf f_j\|_2=1\) and \(0\le q_j\le P_{\max}\). Accordingly, the harvested energy of user \(k\) is \(E_k^{\mathrm{EH}}=\sum_{j=1}^{J}\tau_j\Phi_k\left(q_j|\mathbf h_k^H\mathbf f_j|^2\right)\), where \(\Phi_k(\cdot)\) denotes the nonlinear EH function defined in Section II-B. 
	Using \(t_0=\sum_{j=1}^{J}\tau_j+(J-1)\delta_{\mathrm{sw}}\), we eliminate \(t_0\) from the optimization variables. For uplink transmission, the single-RF-chain SC-MiLAC achieves the same reception performance as fully digital MRC. Therefore, the achievable uplink rate of user \(k\) is given by
	\begin{flalign}
		R_k^{\mathrm{UL}}
		=W\log_2\left(
		1+\frac{p_k\|\mathbf h_k\|^2}
		{\sigma_{\mathrm{BS}}^2}
		\right). \label{eq30}
	\end{flalign}
	
	The offloaded task size satisfies \(0\le \ell_k\le t_kR_k^{\mathrm{UL}}\), where \(t_k\) and \(p_k\) denote the uplink transmission time and transmit power, respectively. Introduce \(z_k\in\{0,1\}\) to indicate whether user \(k\) is scheduled for task offloading, with \(\ell_k\le D_kz_k\).
	
	Following the prescribed TDMA order, the completion time of user \(k\)'s uplink transmission is \(u_k= t_0+\sum_{i=1}^{k}\left(t_i+\delta_{\mathrm{sw}}z_i\right)\). The local computing completion time is \(T_k^{\mathrm{loc}}
	=t_0+
	\frac{c_k(D_k-\ell_k)}
	{f_k^{\mathrm{loc}}}\),
	where \(f_k^{\mathrm{loc}}\) is the local CPU frequency. For an offloading user, the edge computing completion time is \(T_k^{\mathrm{edge}}
	=u_k+\frac{c_k\ell_k}{f_k^{\mathrm{edge}}}\).
	
	Since local computing and task offloading proceed in parallel during the WIT phase, the task completion time is determined by the later of the two execution paths. Introduce \(T\) as an upper bound on the maximum task completion latency. Retaining the edge CPU allocation variables, the edge computing constraint becomes \(c_k\ell_k\le f_k^{\mathrm{edge}}(T-u_k)\), with \(\sum_{k=1}^{K}f_k^{\mathrm{edge}}\le F_{\max}\) and \(0\le f_k^{\mathrm{edge}}\le F_{\max}z_k\). We impose \(T\ge u_K\) to ensure nonnegative edge computing durations. The optimization problem is formulated as
	\begin{subequations} \label{p1}
		\begin{align}
			&\min_{\substack{
					\{\mathbf f_j,q_j,\tau_j\}_{j=1}^{J},
					\{\ell_k,t_k,p_k,f_k^{\mathrm{loc}},f_k^{\mathrm{edge}},z_k\}_{k=1}^{K},T
			}} T \label{p1-a}\\
			\mathrm{s.t.}~
			&t_0+\frac{c_k(D_k-\ell_k)}
			{f_k^{\mathrm{loc}}}\le T, ~\forall k,\\ 
			&c_k\ell_k \leq f_k^{\rm edge}(T-u_k),~\forall k,\label{p1-b}\\
			&\sum\nolimits_{k=1}^{K}f_k^{\mathrm{edge}}\le F_{\max},~u_K \le T, \label{p1-c}\\
			&0\le\ell_k\le D_kz_k,~\ell_k\le t_kR_k^{\mathrm{UL}},~ \forall k, \label{p1-d}\\
			&E_k^{\mathrm{loc}}+E_k^{\mathrm{UL}}
			\le E_k^{\mathrm{EH}},~ \forall k, \label{p1-e}\\
			&t_k\ge 0,~ 0\le p_k\le p_{k,\max},~ \forall k,\\
			&0<f_k^{\mathrm{loc}}\le f_{k,\max}^{\mathrm{loc}},~0\le f_k^{\mathrm{edge}}\le F_{\max}z_k,~ \forall k,\\
			&z_k\in\{0,1\},~ \forall k,\\
			&\tau_j \ge 0,~0\le q_j\le P_{\max},~
			\|\mathbf f_j\|_2=1,~ \forall j. \label{p1-f}
		\end{align}
	\end{subequations}
	
	The constraints \eqref{p1-b}-\eqref{p1-f} impose the task completion, computing resource allocation, task offloading, energy causality, and hardware feasibility requirements, which have been introduced in the preceding sections and are not repeated here. Problem \eqref{p1} is nonconvex due to the nonlinear EH model, the coupling among optimization variables, the unit-norm beamforming constraint, and the binary offloading decisions. To solve this problem, we develop an alternating optimization algorithm that exploits the structure of each variable block.
	
	\subsection{WPT Beamforming Optimization}
	To simplify the beamforming optimization, we define the power-weighted beamforming vector $\mathbf w_j=\sqrt{q_j}\mathbf f_j$, such that $q_j|\mathbf h_k^H\mathbf f_j|^2=|\mathbf h_k^H\mathbf w_j|^2$ and $\|\mathbf w_j\|_2^2\leq P_{\max}$. For fixed time and resource allocation variables, the energy requirement of user \(k\) is given by \(E_k^{\mathrm{req}}=E_k^{\mathrm{loc}}+E_k^{\mathrm{UL}}\). Since the WPT beamforming vectors do not explicitly affect the objective \(T\), we maximize the minimum normalized harvested energy as an auxiliary objective for the beamforming update. The beamforming subproblem is formulated as
	\begin{subequations} \label{p2}
		\begin{align}
			&\max_{\{\mathbf w_j\}_{j=1}^J,\rho}\quad
			\rho \label{p2-a}\\
			\mathrm{s.t.}~
			&\sum\nolimits_{j=1}^{J}\tau_j
			\Phi_k\left(|\mathbf h_k^H\mathbf w_j|^2\right)
			\ge \rho E_k^{\mathrm{req}}, ~\forall k, \label{p2-b}\\
			&\|\mathbf w_j\|_2^2\le P_{\max}, ~\forall j, ~\rho \ge 1, \label{p2-c}
		\end{align}
	\end{subequations}
	where \(\rho\) denotes the minimum normalized harvested energy among users. Problem \eqref{p2} is nonconvex due to the nonlinear EH constraints. To solve it, we develop an inverse-Sigmoid-based DC-SCA method. 
	
	First, introduce an auxiliary variable \(\eta_{k,j}\) representing the guaranteed harvested DC power of user \(k\) during WPT subinterval \(j\). The EH constraints \eqref{p2-b} can be equivalently rewritten as \(\sum_{j=1}^{J}\tau_j\eta_{k,j}
	\ge \rho E_k^{\mathrm{req}}\) and \(|\mathbf h_k^H\mathbf w_j|^2
	\ge \Phi_k^{-1}(\eta_{k,j})\), where \(0\le\eta_{k,j}\le\eta_{k,\max}\), with \(\eta_{k,\max}=\Phi_k(P_{\max}\|\mathbf h_k\|_2^2)\).
	
	Define \(\zeta_k(\eta)=\Omega_k+(1-\Omega_k)\eta/M_k\). The inverse of the nonlinear EH function can be expressed as 
	\begin{flalign}
		\Phi_k^{-1}(\eta)
		=b_k+\frac{1}{a_k}
		\left[
		\ln \zeta_k(\eta)
		-\ln\left(1-\zeta_k(\eta)\right)
		\right].
	\end{flalign} 
	Since both \(-\ln \zeta_k(\eta)\) and \(-\ln(1-\zeta_k(\eta))\) are convex, the inverse EH function admits a DC representation. Accordingly, we construct a convex upper bound by linearizing the concave term \(\ln \zeta_k(\eta)\). At the \(r\)-th SCA iteration, define \(\zeta_{k,j}^{(r)}=\zeta_k(\eta_{k,j}^{(r)})\). By the concavity of the logarithmic function, we have 
	\begin{flalign}
		\ln \zeta_k(\eta)
		\le
		\ln \zeta_{k,j}^{(r)}
		+\frac{\zeta_k(\eta)-\zeta_{k,j}^{(r)}}{\zeta_{k,j}^{(r)}}.
	\end{flalign}
	This yields the convex upper bound
	\begin{flalign}
		\hat{g}_{k,j}^{(r)}(\eta)
		= b_k + \frac{1}{a_k}
		\left[
		\ln \zeta_{k,j}^{(r)}
		+ \frac{\zeta_k(\eta)-\zeta_{k,j}^{(r)}}{\zeta_{k,j}^{(r)}}
		-\ln\left(1-\zeta_k(\eta)\right)
		\right].
	\end{flalign} 
	This bound satisfies \(\Phi_k^{-1}(\eta)\le\hat g_{k,j}^{(r)}(\eta)\), with equality and matching first-order derivatives at \(\eta=\eta_{k,j}^{(r)}\).
	
	Next, let \(\mathbf H_k=\mathbf h_k\mathbf h_k^H\). Since \(\mathbf H_k\succeq\mathbf0\), the received RF power admits the global affine lower bound 
	\begin{flalign}
		|\mathbf{h}_k^H \mathbf{w}_j|^2
		&\geq 2\operatorname{Re}\left\{
		\left(\mathbf{w}_j^{(r)}\right)^H
		\mathbf{H}_k \mathbf{w}_j
		\right\} - \left|\mathbf{h}_k^H \mathbf{w}_j^{(r)}\right|^2 \notag\\
		&\triangleq \underline P_{k,j}^{(r)}(\mathbf{w}_j).
	\end{flalign}
	
	Accordingly, the nonconvex RF power constraint on each \(\eta_{k,j}\) can be conservatively approximated by \(\underline P_{k,j}^{(r)}(\mathbf w_j)
	\ge
	\hat g_{k,j}^{(r)}(\eta_{k,j})\). Since the left-hand side is affine and the right-hand side is convex, this constraint is convex. Consequently, the beamforming subproblem at iteration \(r\) is formulated as
	\begin{subequations} \label{p3}
		\begin{align}
			& \max_{\{\mathbf{w}_j,\eta_{k,j}\},\rho} ~ \rho \\
			\mathrm{s.t.}~ &\sum\nolimits_{j=1}^{J}\tau_j\eta_{k,j}
			\geq \rho E_k^{\mathrm{req}},~\forall k, \\
			& \underline P_{k,j}^{(r)}(\mathbf{w}_j)
			\geq \hat{g}_{k,j}^{(r)}(\eta_{k,j}),~\forall k,j, \\
			& 0\leq\eta_{k,j}\leq\eta_{k,\max},~\forall k,j,\\
			& \|\mathbf{w}_j\|_2^2\leq P_{\max},~\forall j,~ \rho\geq 1.
		\end{align}
	\end{subequations}
	
	Problem \eqref{p3} is convex and can be efficiently solved using standard convex optimization tools \cite{boyd2004convex}.

	\subsection{Joint Time and Resource Allocation}
	For given WPT beamforming vectors \(\{\mathbf w_j\}\), the harvested DC power \(P_{k,j}^{\mathrm{DC}}=\Phi_k(|\mathbf h_k^H\mathbf w_j|^2)\) becomes constant. We jointly optimize the WPT durations, uplink transmission resources, task offloading decisions, and computing resources to minimize the maximum task completion latency.
	
	Define \(e_k=p_kt_k\) as the uplink transmission energy of user \(k\). By exploiting the single-RF-chain receive beamforming result in Section III-D, the offloading constraint can be equivalently rewritten as 
	\begin{flalign}
		\ell_k\le Wt_k\log_2\left(
		1+\frac{e_k\|\mathbf h_k\|_2^2}
		{t_k\sigma_{\mathrm{BS}}^2}
		\right),~\forall k. \label{eq34}
	\end{flalign}
	
	The right-hand side of \eqref{eq34} is the perspective of a concave logarithmic function. Hence, \eqref{eq34} is a convex constraint. Meanwhile, the uplink energy consumption and transmit-power constraint become \(E_k^{\mathrm{UL}}=e_k+P_{k,\mathrm{cir}}^{\mathrm{UL}}t_k\) and \(0\le e_k\le p_{k,\max}t_k\), respectively.
	
	Next, we use the local computing duration \(t_k^{\mathrm{loc}}\) as an optimization variable and eliminate \(f_k^{\mathrm{loc}}\) by substituting \(f_k^{\mathrm{loc}}=c_k(D_k-\ell_k)/t_k^{\mathrm{loc}}\). The local computing energy is then expressed as
	\begin{flalign}
		E_k^{\mathrm{loc}}
		=
		\frac{\xi_k c_k^3(D_k-\ell_k)^3}
		{(t_k^{\mathrm{loc}})^2}
		+P_{k,\mathrm{cir}}^{\mathrm{CPU}}t_k^{\mathrm{loc}}. \label{eq35}
	\end{flalign}
	
	Since the first term in \eqref{eq35} is the perspective of a convex cubic function, \(E_k^{\mathrm{loc}}\) is jointly convex in \((\ell_k,t_k^{\mathrm{loc}})\). The local CPU frequency and task completion constraints are equivalently transformed into
	\begin{flalign}
		c_k(D_k-\ell_k) \le f_{k,\max}^{\mathrm{loc}}t_k^{\mathrm{loc}},~
		t_0+t_k^{\mathrm{loc}}\le T,~\forall k. \label{eq36}
	\end{flalign}
	Here, the local computing energy and duration are defined as zero when \(\ell_k=D_k\).
	
	To handle the binary offloading decisions, we equivalently express \(z_k\in\{0,1\}\) as \(0\le z_k\le1\) and \(z_k-z_k^2=0\). Following the penalty method, the equality is incorporated into the objective through \(\chi\sum_{k=1}^{K}(z_k-z_k^2)\), where \(\chi>0\) is the penalty factor. At iteration \(r\), the concave penalty term admits the affine upper bound
	\begin{flalign}
		z_k-z_k^2
		\leq (1-2z_k^{(r)})z_k+(z_k^{(r)})^2
		\triangleq\hat\varphi_k^{(r)}(z_k). \label{eq37}
	\end{flalign}
	
	The remaining nonconvexity arises from the edge computing constraint \(c_k\ell_k\le f_k^{\mathrm{edge}}(T-u_k)\). Define \(d_k=T-u_k\), \(x_k=f_k^{\mathrm{edge}}/F_{\max}\), and \(y_k=d_k/t_{\mathrm{ref}}\), where \(t_{\mathrm{ref}}>0\) is a time normalization constant. Let \(s_k^{(r)}=x_k^{(r)}+y_k^{(r)}\). The concave lower bound of \(f_k^{\mathrm{edge}}d_k\) at iteration \(r\) is given by
	\begin{flalign}
		f_k^{\mathrm{edge}}d_k
		&\geq F_{\max}t_{\mathrm{ref}}\left[
		\frac{s_k^{(r)}(x_k+y_k)}{2}
		-\frac{(s_k^{(r)})^2}{4}-\frac{(x_k-y_k)^2}{4}
		\right] \notag\\
		&\triangleq
		\underline G_k^{(r)}(f_k^{\mathrm{edge}},d_k). \label{eq38}
	\end{flalign}
	
	Therefore, the edge computing constraint can be conservatively approximated by
	\begin{flalign}
		c_k\ell_k
		\leq\underline G_k^{(r)}(f_k^{\mathrm{edge}},T-u_k),~\forall k. \label{eq39}
	\end{flalign}
	
	Based on the above transformations, the joint time and resource allocation subproblem at iteration \(r\) is formulated as
	\begin{subequations} \label{p4}
		\begin{align}
			& \min_{\boldsymbol{\tau},\boldsymbol{\ell},
				\mathbf{t},\mathbf{e},\mathbf{t}^{\mathrm{loc}},\mathbf{f}^{\mathrm{edge}},\mathbf{z},T}~ T+\chi\sum_{k=1}^{K}\hat\varphi_k^{(r)}(z_k) \\
			\mathrm{s.t.}~& E_k^{\mathrm{loc}}+e_k+P_{k,\mathrm{cir}}^{\mathrm{UL}}t_k
			\leq \sum\nolimits_{j=1}^{J}\tau_j P_{k,j}^{\mathrm{DC}},~\forall k,\\
			& 0\leq\ell_k\leq D_kz_k,~0\leq e_k\leq p_{k,\max}t_k,~t_k \geq 0,\notag\\
			& t_k^{\mathrm{loc}}\geq 0,~0\leq z_k\leq1,~0\leq f_k^{\mathrm{edge}}\leq F_{\max}z_k,~\forall k,\\
			& \sum\nolimits_{k=1}^{K}f_k^{\mathrm{edge}}\leq F_{\max},~T\geq u_K,~\tau_j\geq 0,~ \forall j,\\
			& \eqref{eq34},~\eqref{eq36},~\eqref{eq39}. \notag
		\end{align}
	\end{subequations}
	
	For a fixed penalty factor \(\chi\), problem \eqref{p4} is convex and can be solved using standard convex optimization tools \cite{boyd2004convex}. The penalty factor is gradually increased until the binary constraint violation \(\max_k(z_k-z_k^2)\) meets a prescribed tolerance.
	The original variables are recovered as \(p_k=e_k/t_k\) and \(f_k^{\mathrm{loc}}=c_k(D_k-\ell_k)/t_k^{\mathrm{loc}}\) for nonzero corresponding workloads. The edge CPU frequencies \(f_k^{\mathrm{edge}}\) are directly optimized.

	\subsection{Overall Algorithm and Complexity Analysis}
	\begin{algorithm}[!t]
		\caption{PBDC-SCA Algorithm for Solving Problem \eqref{p1}}
		\label{alg:overall}
		\begin{algorithmic}[1]
			\STATE Initialize feasible \(\{\mathbf w_j^{(0)}\}\), \(\boldsymbol\tau^{(0)}\), \(\boldsymbol\ell^{(0)}\), \(\mathbf t^{(0)}\), \(\mathbf e^{(0)}\), \((\mathbf t^{\mathrm{loc}})^{(0)}\), \((\mathbf f^{\mathrm{edge}})^{(0)}\), \(\mathbf z^{(0)}\), and \(T^{(0)}\). Set \(\chi^{(0)}>0\), \(\omega>1\), \(\epsilon_{\mathrm{SCA}}>0\), \(\epsilon_{\mathrm{AO}}>0\), \(\epsilon_{\mathrm{bin}}>0\), and \(s=0\).
			\REPEAT
			\STATE Set \(r=0\).
			\REPEAT
			\STATE \(\eta_{k,j}^{(r)}=\Phi_k(|\mathbf h_k^H\mathbf w_j^{(r)}|^2)\), construct \(\hat g_{k,j}^{(r)}\) and \(\underline P_{k,j}^{(r)}\).
			\STATE Update \(\{\mathbf w_j^{(r+1)}\}\), \(\{\eta_{k,j}^{(r+1)}\}\), \(\rho^{(r+1)}\) by solving \eqref{p3}.
			\STATE Update \(P_{k,j}^{\mathrm{DC}}=\Phi_k(|\mathbf h_k^H\mathbf w_j^{(r+1)}|^2)\), \(d_k^{(r)}=T^{(r)}-u_k^{(r)}\), construct \(\hat\varphi_k^{(r)}\) and \(\underline G_k^{(r)}\).
			\STATE Update \(\boldsymbol\tau^{(r+1)}\), \(\boldsymbol\ell^{(r+1)}\), \(\mathbf t^{(r+1)}\), \(\mathbf e^{(r+1)}\), \((\mathbf t^{\mathrm{loc}})^{(r+1)}\), \((\mathbf f^{\mathrm{edge}})^{(r+1)}\), \(\mathbf z^{(r+1)}\), \(T^{(r+1)}\) by solving \eqref{p4}.
			\STATE Set \(r\leftarrow r+1\).
			\UNTIL{\(|\rho^{(r)}-\rho^{(r-1)}|\le\epsilon_{\mathrm{SCA}}\) and \(|\mathcal L_{\chi}^{(r)}-\mathcal L_{\chi}^{(r-1)}|\le\epsilon_{\mathrm{AO}}\)}
			\IF{\(\max_k(z_k-z_k^2)>\epsilon_{\mathrm{bin}}\)}
			\STATE Set \(\chi^{(s+1)}=\omega\chi^{(s)}\), \(s\leftarrow s+1\).
			\ENDIF
			\UNTIL{\(\max_k(z_k-z_k^2)\le\epsilon_{\mathrm{bin}}\)}
			\STATE Set \(z_k\leftarrow\operatorname{round}(z_k)\) and reoptimize the continuous variables with fixed \(\{z_k\}\) and \(\chi=0\).
		\end{algorithmic}
	\end{algorithm}
	The proposed penalty-based alternating DC-SCA algorithm is summarized in Algorithm \ref{alg:overall}. For a fixed \(\chi\), the WPT beamforming vectors are updated by solving \eqref{p3} once, followed by updating the time and resource allocation variables, including the offloading indicators \(\{z_k\}\), by solving \eqref{p4} once. These two steps are alternately performed with updated SCA approximations until the convergence criteria are satisfied. The penalty factor is then increased as \(\chi\leftarrow\omega\chi\), where \(\omega>1\), until the binary constraint violation \(\max_k(z_k-z_k^2)\) falls below a prescribed tolerance \(\epsilon_{\mathrm{bin}}\).
	
	For a fixed penalty factor \(\chi\), denote the penalized objective by \(\mathcal L_\chi=T+\chi\sum_{k=1}^{K}(z_k-z_k^2)\). The beamforming update preserves feasibility without changing \(\mathcal L_\chi\). In the resource allocation step, the affine upper bound in \eqref{eq37} is tight at the current iterate, and the lower bound in \eqref{eq38} preserves the original edge computing constraints. Hence, each AO-SCA iteration satisfies \(\mathcal L_\chi^{(r+1)}\le\mathcal L_\chi^{(r)}\).
	Since \(\mathcal L_\chi\ge0\), its objective-value sequence converges for each fixed \(\chi\). Once the binary constraint violation falls below \(\epsilon_{\mathrm{bin}}\), the offloading indicators are rounded to binary values, and the continuous variables are reoptimized with the offloading decisions fixed.
	Let \(I_{\mathrm{P}}\) and \(I_{\mathrm{AO}}\) denote the numbers of penalty and AO-SCA iterations, respectively. Problems \eqref{p3} and \eqref{p4} contain \(n_{\mathrm{BF}}=2NJ+KJ+1\) and \(n_{\mathrm{RA}}=J+6K+1\) real optimization variables, respectively, excluding auxiliary variables introduced by the convex solver. Their numbers of constraints scale as \(m_{\mathrm{BF}}=\mathcal O(KJ+K+J)\) and \(m_{\mathrm{RA}}=\mathcal O(K+J)\). Using a generic dense interior-point implementation, the overall computational complexity can be estimated as \(\mathcal O\left(I_{\mathrm{P}}I_{\mathrm{AO}}\big[
	\sqrt{m_{\mathrm{BF}}}\,n_{\mathrm{BF}}^3
	+\sqrt{m_{\mathrm{RA}}}\,n_{\mathrm{RA}}^3
	\big]\ln\frac{1}{\varepsilon}\right)\),
	where \(\varepsilon\) denotes the convex solver accuracy.
	
	\section{Numerical Results}
	
	In this section, we provide numerical results to evaluate the performance of the proposed SC-MiLAC-enabled WP-MEC system. Unless otherwise stated, the number of BS antennas and users are set to \(N=8\) and \(K=4\), respectively, with \(J=2\) WPT beam configurations. The BS employs a ULA with an inter-antenna spacing of \(\lambda/2\), operating at a carrier frequency of \(f_c=2.4~{\rm GHz}\) and a bandwidth of \(W=0.5~{\rm MHz}\). The BS antenna array is located at a height of \(h_{\rm BS}=3~{\rm m}\), while the users are placed on the ground at a horizontal distance of 10 m from the BS, with their angular positions equally spaced over \([-40^\circ,40^\circ]\). The BS-user channels are modeled as Rician fading channels. The Rician factor is set to \(3~{\rm dB}\). The reference channel power gain is set to \(\beta_0=-30~{\rm dB}\) at \(1~{\rm m}\), with a path-loss exponent of \(\alpha=2.2\). The receiver noise figure is set to \(7~\rm dB\), corresponding to a noise power of approximately \(-110~\rm dBm\) over the considered bandwidth. The maximum BS transmit power and user transmit power are set to \(P_{\max}=35~\rm dBm\) and \(p_{k,\max}=10~\rm dBm\), respectively. Each user has a computation task of \(D_k=20~\rm kbits\), with \(c_k=300\) CPU cycles per bit. The maximum local CPU frequency and the total MEC server computing capacity are set to \(f_{k,\max}^{\rm loc}=0.5~\rm GHz\) and \(F_{\max}=1~\rm GHz\), respectively. The parameters of the nonlinear EH model are set to \(M_k=24~\rm mW\), \(a_k=150\) W\(^{-1}\), and \(b_k=14~\rm mW\). The local computing and uplink circuit powers are set to \(P_{k,\rm cir}^{\rm CPU}=10\) \(\mu \rm W\) and \(P_{k,\rm cir}^{\rm UL}=100\) \(\mu \rm W\), respectively. The switching delay is set to \(\delta_{\rm sw}=0.1~\rm ms\) for the practical dynamic SC-MiLAC scheme, while the other benchmark schemes are assumed to have negligible switching delays. 
	For performance comparison, six benchmark schemes are considered as follows.
	\begin{itemize}
		\item \textbf{Ideal Dynamic SC-MiLAC}: An SC-MiLAC architecture with one RF chain and $N$ antennas with \(\delta_{\rm sw} = 0~\rm ms\).
		\item \textbf{Dynamic SC-MiLAC scheme}: The same architecture as the ideal dynamic SC-MiLAC, but \(\delta_{\rm sw} = 0.1~\rm ms\).
		\item \textbf{Fully digital scheme}: A fully digital architecture with $N$ RF chains, adopting the same dynamic configuration as the SC-MiLAC schemes, without switching delay.
		\item \textbf{Static SC-MiLAC scheme}: A fixed MiLAC configuration is shared throughout the entire WPT and WIT phases.
		\item \textbf{Dynamic ABF scheme}: An analog beamforming architecture with one RF chain and $N$ antennas, allowing dynamic beam configuration without switching delay.
		\item \textbf{Conventional ABF scheme}: An analog beamforming architecture with one RF chain and $N$ antennas, using a common phase-shifter configuration throughout the WPT phase while allowing configuration switching during the WIT phase, without switching delay \cite{7400949}.
	\end{itemize}
	
	\begin{figure}[!t]
		\centering
		\includegraphics[width=0.44\textwidth]{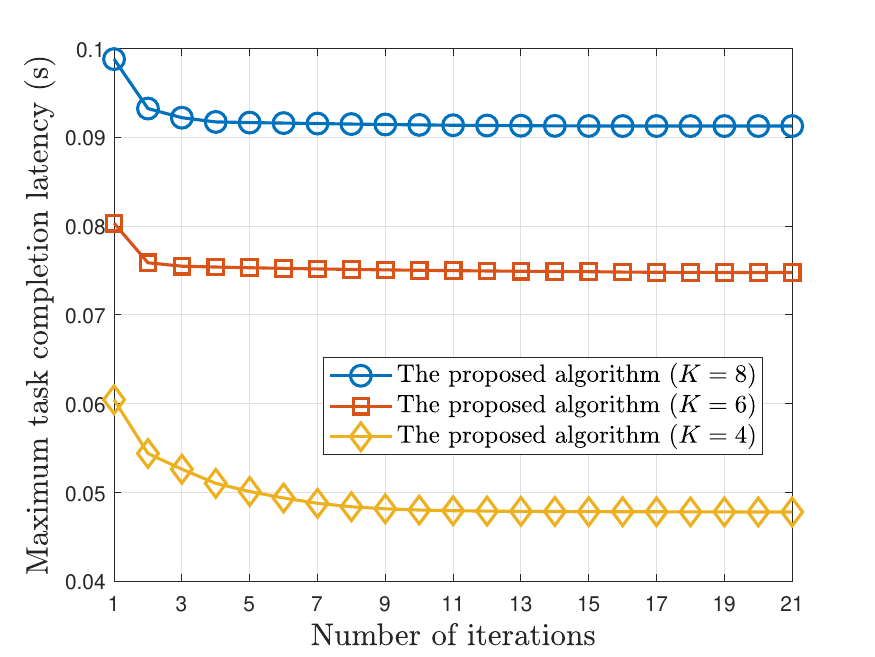}
		\caption{Convergence of the proposed PBDC-SCA algorithm.}
		\label{iter}
	\end{figure}
	Fig.~\ref{iter} illustrates the convergence behavior of the proposed PBDC-SCA algorithm under different numbers of users. It can be observed that the maximum task completion latency decreases rapidly during the initial iterations and gradually converges to a stable value. In particular, the algorithm achieves convergence within approximately 10 iterations for all considered cases, demonstrating its favorable convergence performance. Moreover, the maximum task completion latency increases with the number of users, as more users compete for the limited WPT energy, uplink transmission resources, and MEC computing capacity.
	
	\begin{figure}[!t]
		\centering
		\includegraphics[width=0.44\textwidth]{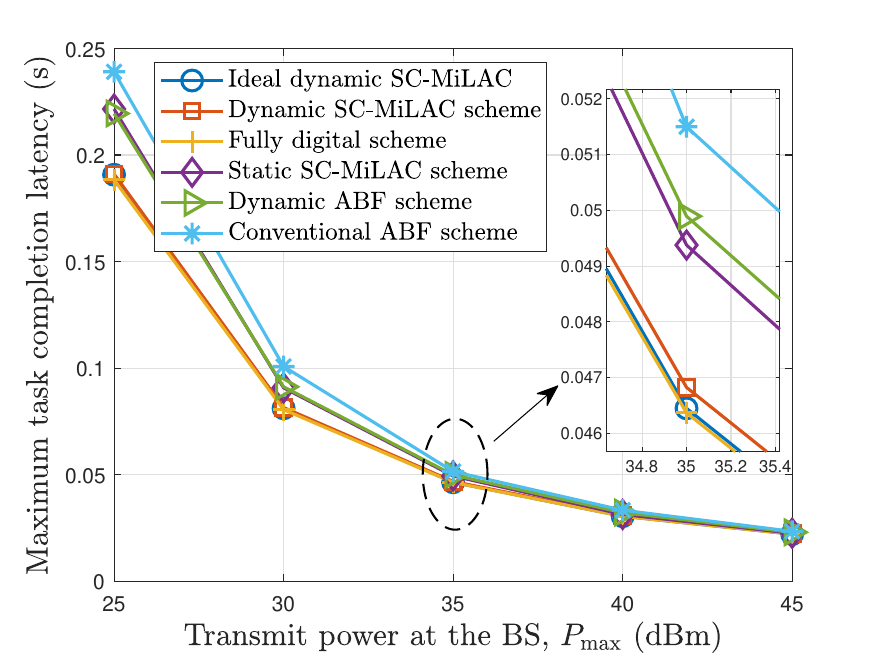}
		\caption{Maximum task completion latency versus the maximum BS transmit power.}
		\label{Pmax}
	\end{figure}
	Fig.~\ref{Pmax} illustrates the maximum task completion latency versus the maximum BS transmit power $P_{\max}$. It can be observed that the latency decreases for all schemes as $P_{\max}$ increases, since a higher transmit power enables the users to harvest the required energy within a shorter WPT duration, thereby reducing the overall task completion latency. Moreover, the performance gaps among different schemes become smaller as $P_{\max}$ increases, indicating that the impact of the hardware architecture on system performance becomes less pronounced when sufficient transmit power is available. Specifically, the ideal dynamic SC-MiLAC achieves a maximum task completion latency very close to that of the fully digital scheme, while the practical dynamic SC-MiLAC incurs a slight additional latency due to the switching overhead. In contrast, the static SC-MiLAC and ABF schemes generally exhibit higher latency, owing to their more restricted beamforming configurations.
	
	\begin{figure}[!t]
		\centering
		\includegraphics[width=0.44\textwidth]{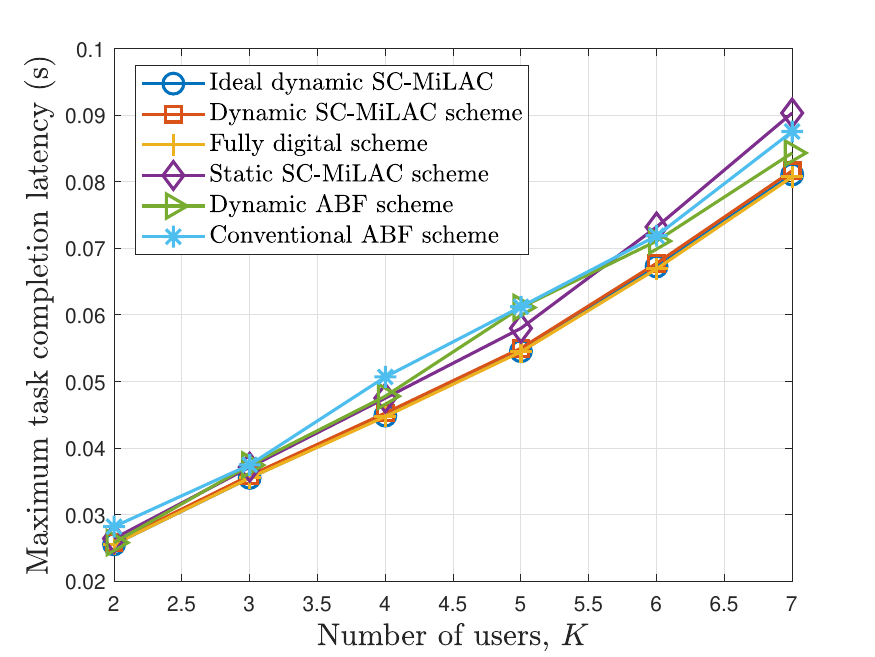}
		\caption{Maximum task completion latency versus the number of users.}
		\label{K}
	\end{figure}
	Fig.~\ref{K} illustrates the maximum task completion latency versus the number of users $K$. It can be observed that the latency increases for all schemes as $K$ increases, since more users need to share the limited energy, uplink transmission, and MEC computing resources. Notably, the performance of the static SC-MiLAC scheme deteriorates more rapidly with increasing $K$. This is because a single fixed analog configuration must serve all users throughout the WPT and WIT phases, making it difficult to accommodate the different channel conditions and transmission requirements of multiple users. As the number of users increases, this limitation becomes more pronounced, resulting in a growing performance gap compared with the dynamic schemes.
	
	\begin{figure*}[!t]
		\centering
		\subfloat[$T$ versus $J$ with $K=4$.]{%
			\includegraphics[width=0.32\linewidth]{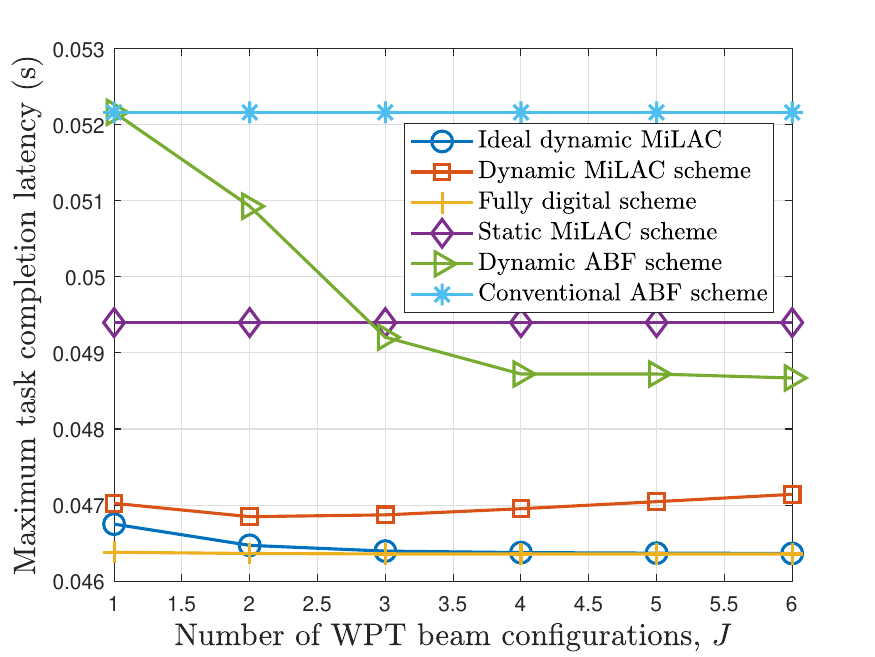}%
			\label{a}}
		\hfill
		\subfloat[$T$ versus $J$ with $K=6$.]{%
			\includegraphics[width=0.32\linewidth]{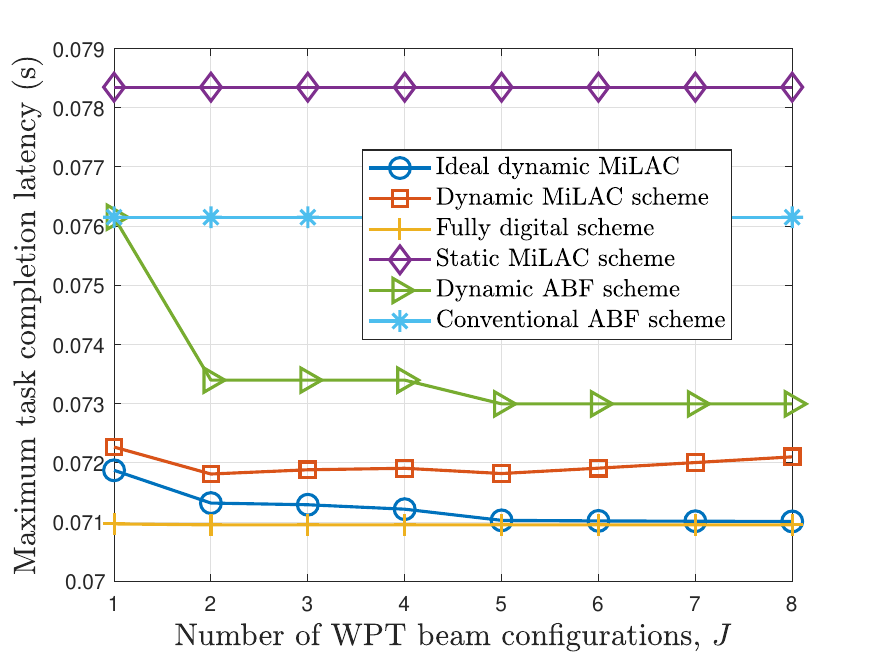}%
			\label{b}}
		\hfill
		\subfloat[$T$ versus $J$ with $K=8$.]{%
			\includegraphics[width=0.32\linewidth]{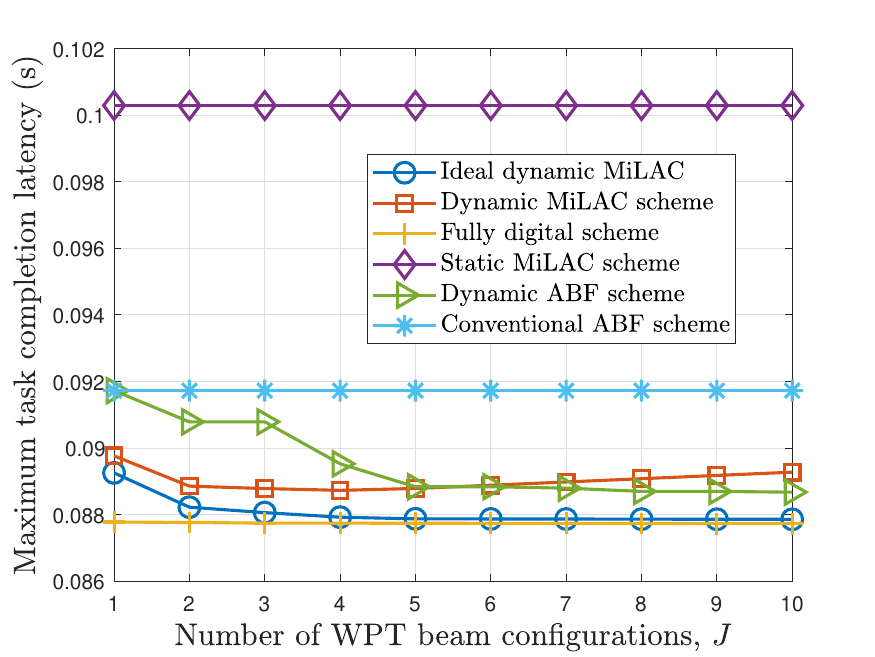}%
			\label{c}}
		\caption{Maximum task completion latency versus the number of WPT beam configurations for $K=4$, $6$, and $8$, respectively.}
		\label{J}
	\end{figure*} 
	Fig.~\ref{J} illustrates the maximum task completion latency versus the number of WPT beam configurations $J$ under different numbers of users, i.e., $K=4$, $6$, and $8$. It can be observed that the latency of the ideal dynamic SC-MiLAC and dynamic ABF schemes decreases with increasing $J$ and gradually converges, since more beam configurations provide greater flexibility in adapting energy transmission to different users. In contrast, the practical dynamic SC-MiLAC exhibits a non-monotonic trend, where the latency initially decreases and then slightly increases. This is because increasing $J$ introduces additional switching overhead, which must be balanced against the EH gains brought by more flexible beam configurations. When the additional beamforming gains can no longer compensate for the increased switching delay, the overall task completion latency starts to increase. Meanwhile, the static SC-MiLAC and conventional ABF schemes exhibit nearly constant performance with increasing $J$, as their WPT phases rely on fixed analog configurations. Notably, the performance gap between the static SC-MiLAC and dynamic schemes becomes more pronounced as the number of users increases, since a single fixed analog configuration becomes increasingly inadequate for accommodating the diverse channel conditions of multiple users. 
	
	\begin{figure}[!t]
		\centering
		\includegraphics[width=0.44\textwidth]{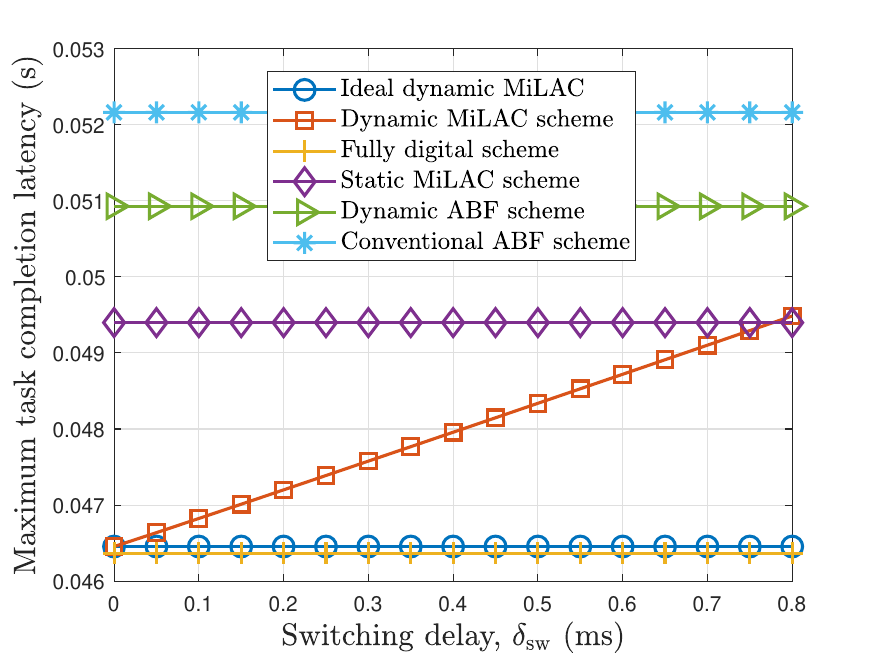}
		\caption{Maximum task completion latency versus the switching delay.}
		\label{delta}
	\end{figure}
	Fig.~\ref{delta} illustrates the maximum task completion latency versus the switching delay $\delta_{\rm sw}$. It can be observed that the latency of the practical dynamic SC-MiLAC scheme increases with $\delta_{\rm sw}$, while the other schemes remain nearly unaffected due to their assumed negligible switching delays. This is because a larger switching delay introduces additional overhead during both the WPT and WIT phases. Specifically, the switching intervals reduce the effective energy transmission time for a given WPT duration, while reconfiguration before uplink offloading further delays task execution. Consequently, more time may be required to satisfy the users' energy demands and complete their computation tasks. As $\delta_{\rm sw}$ increases, the performance advantage of dynamic SC-MiLAC gradually diminishes, eventually approaching that of the static SC-MiLAC scheme. These results highlight the tradeoff between the beamforming flexibility provided by dynamic reconfiguration and the associated hardware switching overhead.
	
	\begin{figure}[!t]
		\centering
		\includegraphics[width=0.44\textwidth]{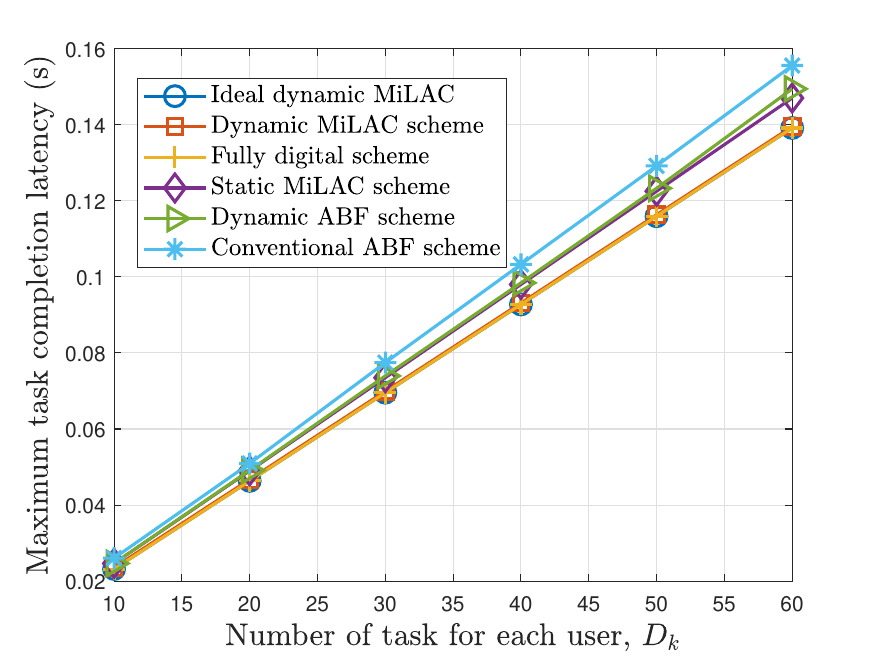}
		\caption{Maximum task completion latency versus the task size per user}
		\label{D}
	\end{figure}
	Fig.~\ref{D} illustrates the maximum task completion latency versus the task size per user $D_k$. As expected, the latency increases for all schemes as $D_k$ increases, since larger tasks require more CPU cycles for local and edge computing, as well as more transmission resources for task offloading. The increased computing and communication demands also lead to higher energy consumption, which may require longer WPT durations to satisfy the users' energy constraints. Moreover, the ideal dynamic SC-MiLAC achieves latency close to that of the fully digital scheme, while the practical dynamic SC-MiLAC maintains comparable performance despite the additional switching overhead. In contrast, the static SC-MiLAC and ABF schemes generally exhibit higher latency, with their performance gaps becoming more pronounced as the task size increases. This is mainly because their more restricted beamforming capabilities limit the flexibility of energy transfer and uplink reception, making it more difficult to accommodate the increasing computation and offloading demands.
	
	\begin{figure}[!t]
		\centering
		\includegraphics[width=0.46\textwidth]{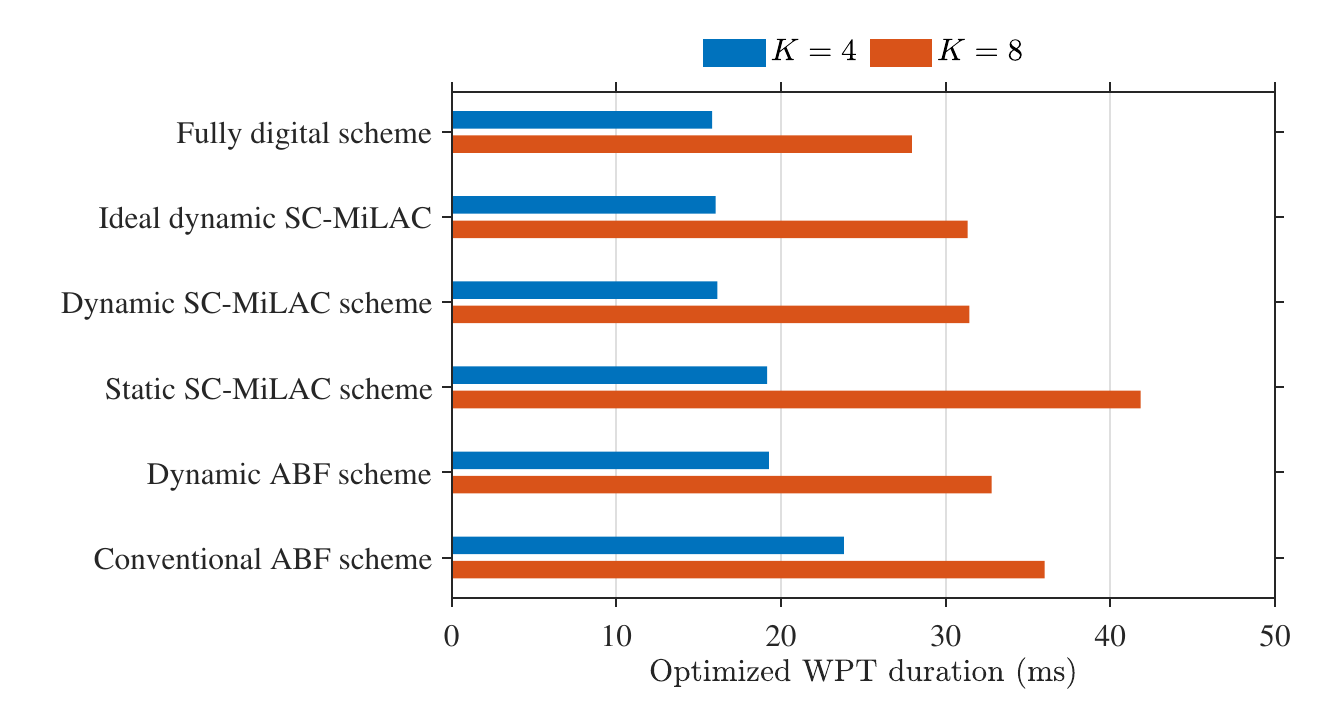}
		\caption{Optimized WPT duration under different schemes.}
		\label{t0}
	\end{figure}
	Fig.~\ref{t0} compares the optimized WPT durations of different schemes for $K=4$ and $K=8$. It can be observed that the WPT duration increases for all schemes as the number of users increases, reflecting the higher aggregate energy demands and increased competition for computing and communication resources. The fully digital scheme achieves the shortest WPT duration, while the ideal dynamic SC-MiLAC exhibits nearly identical performance. The practical dynamic SC-MiLAC requires only a slightly longer WPT duration due to the additional switching overhead. In contrast, the dynamic ABF scheme requires more WPT time because its phase-only beamforming limits the flexibility of energy allocation. The static SC-MiLAC and conventional ABF schemes generally require even longer WPT durations, particularly when $K=8$, since their fixed WPT configurations cannot effectively adapt energy transmission to users with different channel conditions and energy requirements. These results demonstrate that dynamic SC-MiLAC can substantially reduce the required WPT duration by exploiting flexible beam reconfiguration, even with a single RF chain.
	
	\section{Conclusion}
	This paper investigated a single-RF-chain SC-MiLAC-enabled WP-MEC network with dynamic WPT and uplink TDMA, considering nonlinear EH, hardware switching delays, partial task offloading, and limited edge computing resources. We formulated a joint beamforming and resource allocation problem to minimize the maximum task completion latency and developed a PBDC-SCA algorithm to solve it. Theoretical analysis demonstrated that single-RF-chain SC-MiLAC can achieve the same per-user average received RF power as fully digital WPT with at most $K$ beam configurations when switching delays are negligible, while matching fully digital MRC reception under TDMA. Numerical results verified the fast convergence of the proposed algorithm and showed that ideal dynamic SC-MiLAC achieves near-fully-digital latency. With a practical switching delay of 0.1 ms, dynamic SC-MiLAC still outperforms static SC-MiLAC and both ABF benchmarks. These findings demonstrate the effectiveness of dynamic SC-MiLAC in approaching fully digital performance with substantially reduced RF-chain requirements, while highlighting the importance of managing hardware switching overhead.

	\bibliography{Reference}

@ARTICLE{7879258,
  author={Mach, Pavel and Becvar, Zdenek},
  journal={IEEE Commun. Surv. Tutorials}, 
  title={Mobile Edge Computing: A Survey on Architecture and Computation Offloading}, 
  year={2017},
  month={3rd Quart.},
  volume={19},
  number={3},
  pages={1628-1656},
  doi={10.1109/COMST.2017.2682318}}

@ARTICLE{8016573,
  author={Mao, Yuyi and You, Changsheng and Zhang, Jun and Huang, Kaibin and Letaief, Khaled B.},
  journal={IEEE Commun. Surv. Tutorials}, 
  title={A Survey on Mobile Edge Computing: The Communication Perspective}, 
  year={2017},
  month={4th Quart.},
  volume={19},
  number={4},
  pages={2322-2358},
  doi={10.1109/COMST.2017.2745201}}

@ARTICLE{7488250,
  author={Shi, Weisong and Cao, Jie and Zhang, Quan and Li, Youhuizi and Xu, Lanyu},
  journal={IEEE Internet Things J.}, 
  title={Edge Computing: Vision and Challenges}, 
  year={2016},
  month={Oct.},
  volume={3},
  number={5},
  pages={637-646},
  doi={10.1109/JIOT.2016.2579198}}

@ARTICLE{7572018,
  author={Mao, Yuyi and Zhang, Jun and Letaief, Khaled B.},
  journal={IEEE J. Sel. Areas Commun.}, 
  title={Dynamic Computation Offloading for Mobile-Edge Computing With Energy Harvesting Devices}, 
  year={2016},
  month={Dec.},
  volume={34},
  number={12},
  pages={3590-3605},
  doi={10.1109/JSAC.2016.2611964}}

@ARTICLE{7762913,
  author={You, Changsheng and Huang, Kaibin and Chae, Hyukjin and Kim, Byoung-Hoon},
  journal={IEEE Trans. Wireless Commun.}, 
  title={Energy-Efficient Resource Allocation for Mobile-Edge Computation Offloading}, 
  year={2017},
  month={Mar.},
  volume={16},
  number={3},
  pages={1397-1411},
  doi={10.1109/TWC.2016.2633522}}

@ARTICLE{8234686,
  author={Wang, Feng and Xu, Jie and Wang, Xin and Cui, Shuguang},
  journal={IEEE Trans. Wireless Commun.}, 
  title={Joint Offloading and Computing Optimization in Wireless Powered Mobile-Edge Computing Systems}, 
  year={2018},
  month={Mar.},
  volume={17},
  number={3},
  pages={1784-1797},
  doi={10.1109/TWC.2017.2785305}}

@ARTICLE{8334188,
  author={Bi, Suzhi and Zhang, Ying Jun},
  journal={IEEE Trans. Wireless Commun.}, 
  title={Computation Rate Maximization for Wireless Powered Mobile-Edge Computing With Binary Computation Offloading}, 
  year={2018},
  month={Jun.},
  volume={17},
  number={6},
  pages={4177-4190},
  doi={10.1109/TWC.2018.2821664}}

@ARTICLE{8794550,
  author={Zeng, Ming and Nguyen, Nam-Phong and Dobre, Octavia A. and Poor, H. Vincent},
  journal={IEEE Wireless Commun. Lett.}, 
  title={Delay Minimization for {NOMA}-Assisted {MEC} Under Power and Energy Constraints}, 
  year={2019},
  month={Dec.},
  volume={8},
  number={6},
  pages={1657-1661},
  doi={10.1109/LWC.2019.2934453}}

@ARTICLE{7130662,
  author={Sardellitti, Stefania and Scutari, Gesualdo and Barbarossa, Sergio},
  journal={IEEE Trans. Signal Inf. Process. Networks}, 
  title={Joint Optimization of Radio and Computational Resources for Multicell Mobile-Edge Computing}, 
  year={2015},
  month={Jun.},
  volume={1},
  number={2},
  pages={89-103},
  doi={10.1109/TSIPN.2015.2448520}}

@ARTICLE{6954434,
  author={Liu, Liang and Zhang, Rui and Chua, Kee-Chaing},
  journal={IEEE Trans. Commun.}, 
  title={Multi-Antenna Wireless Powered Communication With Energy Beamforming}, 
  year={2014},
  month={Dec.},
  volume={62},
  number={12},
  pages={4349-4361},
  doi={10.1109/TCOMM.2014.2370035}}

@ARTICLE{6489506,
  author={Zhang, Rui and Ho, Chin Keong},
  journal={IEEE Trans. Wireless Commun.}, 
  title={{MIMO} Broadcasting for Simultaneous Wireless Information and Power Transfer}, 
  year={2013},
  month={May},
  volume={12},
  number={5},
  pages={1989-2001},
  doi={10.1109/TWC.2013.031813.120224}}

@ARTICLE{9180255,
  author={Ma, Ganggang and Xu, Jie and Liu, Ya-Feng and Moghadam, Mohammad R. Vedady},
  journal={IEEE Wireless Commun. Lett.}, 
  title={Time-Division Energy Beamforming for Multiuser Wireless Power Transfer With Non-Linear Energy Harvesting}, 
  year={2021},
  month={Jan.},
  volume={10},
  number={1},
  pages={53-57},
  doi={10.1109/LWC.2020.3020324}}

@ARTICLE{7400949,
  author={Heath, Robert W. and Gonz{\'a}lez-Prelcic, Nuria and Rangan, Sundeep and Roh, Wonil and Sayeed, Akbar M.},
  journal={IEEE J. Sel. Top. Signal Process.}, 
  title={An Overview of Signal Processing Techniques for Millimeter Wave {MIMO} Systems}, 
  year={2016},
  month={Apr.},
  volume={10},
  number={3},
  pages={436-453},
  doi={10.1109/JSTSP.2016.2523924}}

@ARTICLE{8932399,
  author={Zhang, Jun and Yu, Xianghao and Letaief, Khaled B.},
  journal={IEEE Open J. Commun. Soc.}, 
  title={Hybrid Beamforming for {5G} and Beyond Millimeter-Wave Systems: A Holistic View}, 
  year={2020},
  volume={1},
  number={},
  pages={77-91},
  doi={10.1109/OJCOMS.2019.2959595}}

@ARTICLE{11281875,
  author={Nerini, Matteo and Clerckx, Bruno},
  journal={IEEE Trans. Signal Process.}, 
  title={Analog Computing for Signal Processing and Communications – {Part I}: Computing With Microwave Networks}, 
  year={2025},
  volume={73},
  number={},
  pages={5183-5197},
  doi={10.1109/TSP.2025.3640931}}

@ARTICLE{11281878,
  author={Nerini, Matteo and Clerckx, Bruno},
  journal={IEEE Trans. Signal Process.}, 
  title={Analog Computing for Signal Processing and Communications – {Part II}: Toward Gigantic {MIMO} Beamforming}, 
  year={2025},
  volume={73},
  number={},
  pages={5198-5212},
  doi={10.1109/TSP.2025.3641053}}

@INPROCEEDINGS{11431737,
  author={Nerini, Matteo and Clerckx, Bruno},
  booktitle={GLOBECOM 2025 - 2025 IEEE Global Communications Conference}, 
  title={Analog Computing with Microwave Networks}, 
  year={2025},
  volume={},
  number={},
  pages={907-912},
  doi={10.1109/GLOBECOM59602.2025.11431737}}

@ARTICLE{11455580,
  author={Nerini, Matteo and Clerckx, Bruno},
  journal={IEEE Trans. Wireless Commun.}, 
  title={{MIMO} Systems Aided by Microwave Linear Analog Computers: Capacity-Achieving Architectures With Reduced Circuit Complexity}, 
  year={2026},
  volume={25},
  number={},
  pages={14597-14610},
  doi={10.1109/TWC.2026.3674747}}

@article{zhang2026beamforming,
  title={Beamforming Design for Stem-Connected Microwave Linear Analog Computer ({MiLAC})-Aided Multiuser {MISO} Downlinks},
  author={Zhang, Yuchen and Wu, Zheyu and Clerckx, Bruno and Al-Naffouri, Tareq Y},
  journal={arXiv preprint arXiv:2606.14499},
  year={2026}
}

@article{zhang2026many,
  title={How Many {RF} Chains Does a Microwave Linear Analog Computer ({MiLAC}) Need to Match the Fully-Digital Cram{\'e}r-Rao Bound?},
  author={Zhang, Yuchen and Ge, Yu and Clerckx, Bruno and Al-Naffouri, Tareq Y},
  journal={arXiv preprint arXiv:2606.23986},
  year={2026}
}

@ARTICLE{10107791,
  author={Li, Baogang and Liao, Jia and Wu, Wenjing and Li, Yonghui},
  journal={IEEE Trans. Mob. Comput.}, 
  title={Intelligent Reflecting Surface Assisted Secure Computation of Wireless Powered {MEC} System}, 
  year={2024},
  month={Apr.},
  volume={23},
  number={4},
  pages={3048-3059},
  doi={10.1109/TMC.2023.3269791}}

@article{park2021latency,
  title={Latency minimization for wireless powered mobile edge computing networks with nonlinear rectifiers},
  author={Park, Junhee and Solanki, Sourabh and Baek, Seunghwan and Lee, Inkyu},
  journal={IEEE Trans. Veh. Technol.},
  volume={70},
  number={8},
  pages={8320--8324},
  month={Aug.},
  year={2021},
  publisher={IEEE}
}

@ARTICLE{9957071,
  author={Xie, Wenwu and Li, Yixuan and Yu, Chao and Wang, Ji and Peng, Xin and Zhu, Peng},
  journal={IEEE Commun. Lett.}, 
  title={Sum Rate Maximization for Self-Sustainable {IRS}-Assisted {UAV} Communications}, 
  month={Feb.},
  year={2023},
  volume={27},
  number={2},
  pages={640-644},
  doi={10.1109/LCOMM.2022.3223876}}

@ARTICLE{11206485,
  author={Li, Yixuan and Xu, Hongbo and Zeng, Ming and Liu, Yuanwei},
  journal={IEEE Wireless Commun. Lett.}, 
  title={Pinching Antenna-Aided Wireless Powered Communication Networks}, 
  month={},
  year={2026},
  volume={15},
  number={},
  pages={255-259},
  doi={10.1109/LWC.2025.3622730}}

@article{akram2026scalable,
  title={A Scalable and Integrated Reconfigurable Intelligent Surface},
  author={Akram, Zaid and Elsayed, Mostafa and Hameed, Hira and Ali, Mirza Shujaat and Kazim, Jalil ur Rehman and Imran, Muhammad Ali and H. Abbasi, Qammer},
  journal={Adv. Electron. Mater.},
  volume={12},
  number={1},
  pages={e00674},
  month={Jan.},
  year={2026},
  publisher={Wiley Online Library}
}

@ARTICLE{11618930,
  author={Qin, Haohao and Gu, Bowen and Li, Dong and Yu, Xianhua and Wang, Liejun and Liu, Yuanwei and Sun, Sumei},
  journal={IEEE Trans. Mob. Comput.}, 
  title={Fairness-Aware Computation Offloading in Wireless-Powered {MEC} Systems With Cooperative Energy Recycling}, 
  year={2026},
  month={Nov.},
  volume={25},
  number={11},
  pages={21330-21344},
  doi={10.1109/TMC.2026.3716312}}

@book{boyd2004convex,
  title={Convex optimization},
  author={Boyd, Stephen and Vandenberghe, Lieven},
  year={2004},
  month={Mar.},
  publisher={Cambridge university press}
}
	\bibliographystyle{IEEEtran}
	
\end{document}